\documentclass[aps,prl,twocolumn,groupedaddress,amsmath,amssymb]{revtex4-1}
\usepackage{graphicx}
\usepackage{float} 
\usepackage{xcolor}

\begin{document}

\title{Turbulence generation through an iterative cascade of the elliptical instability}

\author{Ryan McKeown$^1$, Rodolfo Ostilla-M\'{o}nico$^{2,*}$, Alain Pumir$^3$, Michael P. Brenner$^{1,4}$, and  Shmuel M. Rubinstein$^{1,*}$}
\affiliation{$^1$ School of Engineering and Applied Sciences, Harvard University, Cambridge, MA 02138, USA \\
$^2$ Department of Mechanical Engineering, University of Houston, Houston, TX 77204, USA \\
$^3$ Univ Lyon, ENS de Lyon, Laboratoire de Physique, F-69342 Lyon, France \\
$^4$ Google Research, Mountain View, CA 94043, USA \\
$^*$ Corresponding Authors: rostilla@central.uh.edu, shmuel@seas.harvard.edu }

\begin{abstract}
The essence of turbulent flow is the conveyance of energy through the formation, interaction, and destruction of eddies over a wide range of spatial scales—from the largest scales where energy is injected, down to the smallest scales where it is dissipated through viscosity. 
Currently, there is no mechanistic framework that captures how the interactions of vortices drive this cascade. 
We show that iterations of the elliptical instability, arising from the interactions between counter-rotating vortices, lead to the emergence of turbulence. 
We demonstrate how the nonlinear development of the elliptical instability generates an ordered array of antiparallel secondary filaments. The secondary filaments mutually interact, leading to the formation of even smaller tertiary filaments. 
In experiments and simulations, we observe two and three iterations of this cascade, respectively. Our observations indicate that the elliptical instability could be one of the fundamental mechanisms by which the turbulent cascade develops.
\end{abstract}


\maketitle



Understanding how turbulent flows develop and organize has puzzled scientists and engineers for centuries~\cite{Lavin:2018}. The foundational characterization of turbulent flow began with Reynolds over a century ago~\cite{Reynolds:1895} and was quickly followed by rigorous statistical interpretations of how turbulent flows develop~\cite{Taylor:1935,Richardson:1922,Falkovich:2006}. 
In 1937, Taylor and Green~\cite{Taylor:1937} introduced an initial flow condition which produces a cascade of energy from large to small scales. Subsequently, Kolmogorov postulated that turbulent flows exhibit universal behaviors over many length scales. 
Kolmogorov~\cite{Kolmogorov:1941} predicted that within an inertial subrange, the energy spectrum of a turbulent flow has a universal, self-similar form wherein the energy scales as the inverse $5/3$ power of the wavenumber  $k$.  Kolmogorov's energy cascade has been observed in a plethora of experimental systems and numerical simulations, from wind tunnels to river beds, see e.g. Fig.13 of~\cite{Chapman:1979}.

The efficient conveyance of energy from the large scales, where it is injected, to the small scales, where it is dissipated, is at the heart of how complex, three dimensional flows are maintained. It is thus critical to understand how small-scale flow structures are formed and maintained at high Reynolds numbers. In spite of major progress in providing an effective statistical description of turbulent flows~\cite{Pope:2000}, our understanding of the mechanisms by which interactions between eddies are mediated remains limited. In fact, the explanations of how this occurs in real-space are often abstract and ``poetic''~\cite{Richardson:1922, Betchov:1976, Keylock:2016}.

The temporal development of the turbulent cascade remains one of the most intriguing mysteries in fluid mechanics. 
In particular, it is not well understood what specific mechanisms lead to the development of large velocity gradients in turbulent flows.  
These large velocity gradients, which derive from the interactions of turbulent eddies, amplify the kinetic energy dissipation rate, $\epsilon$, in a manner that is independent of the fluid viscosity in the high-Reynolds number limit~\cite{Falkovich:2006, Frisch:1995}.
This implies the existence of an inertial mechanism by which vortices locally interact to convey energy across scales such that the statistical properties of the energy cascade develop in accordance with the scaling laws established by Kolmogorov.
Note that the limit to how fast an initially regular flow may produce extremely large velocity gradients is also a celebrated mathematical problem~\cite{Fefferman:2006}.
It is therefore of great interest to look directly for elementary flow configurations of interacting vortices that begin smooth and rapidly develop into turbulence. 
This approach was implemented by Lundgren~\cite{Lundgren:1982}, who analytically examined the breakdown of a single vortex under axial strain, bursting into an ensemble of helical vortex bundles. While this configuration has been observed to lead to the development of a turbulent flow, it requires the presence of a particular, large-scale strain configuration acting on an isolated vortex~\cite{Cuypers:2003}. 
We implement a more general flow configuration that typifies the fundamental components of the turbulent cascade: the collision of two identical vortices.
Recent numerical and experimental works demonstrate that the breakdown of colliding vortex rings at intermediate Reynolds numbers gives rise to small-scale flow structures~\cite{McKeown:2018}, mediated by the iterative flattening and splitting of the vortex cores to smaller and smaller filaments~\cite{Brenner:2016,PumirSig:1987}.

Here, we revisit the emergence of a turbulent burst of fine-scale flow structures that results from the violent, head-on collision of two coherent vortex rings~\cite{Lim:1992,McKeown:2018}. 
This classical configuration is a unique model system for probing the development of turbulence without any rigid boundaries or large-scale constraints.
We show that for high Reynolds numbers, the violent breakdown of the colliding vortex rings into a turbulent ``soup'' of interacting vortices is mediated by the elliptical instability. 
During the late-stage, nonlinear development of the elliptical instability, an ordered array of antiparallel secondary vortex filaments emerges perpendicular to the collision plane.
Locally, these pairs of counter-rotating secondary filaments spawn another generation of tertiary vortex filaments, resulting in the expeditious formation of a hierarchy of vortices over many scales. Our numerical simulations show that at this stage of the breakdown, the interacting tangle of vortices reaches a turbulent state, such that the energy spectrum of the flow exhibits Kolmogorov scaling. We observe both experimentally and numerically how the elliptical instability precipitates the onset of turbulence, generating and maintaining the means by which the energy of the flow cascades from large to small scales. 

\begin{figure}[ht!]
\centering
\includegraphics[width=\linewidth]{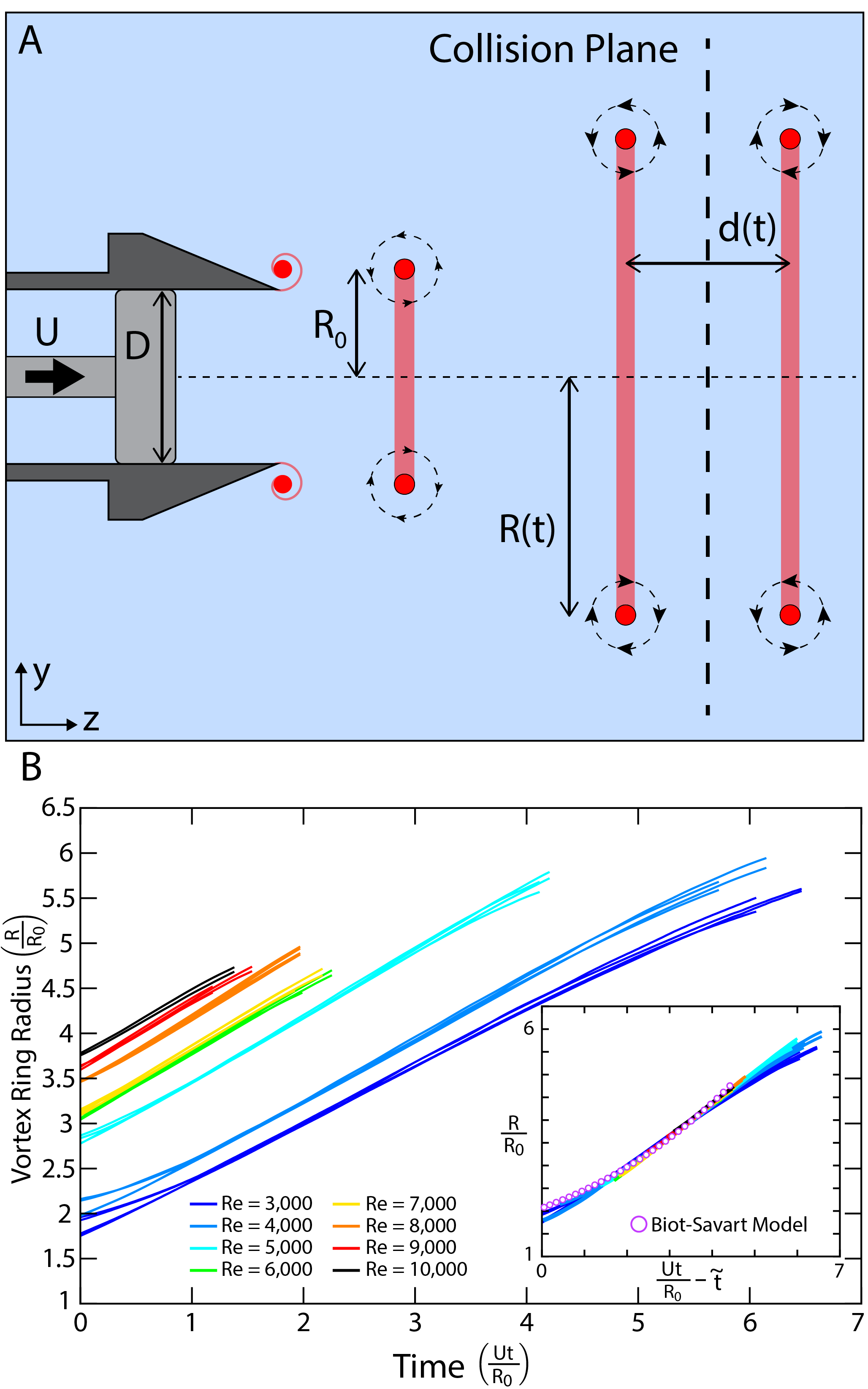}
\caption{
Vortex ring collisions. (A) Schematic side-view showing the formation and collision of dyed vortex rings in experiments. Fluorescent dye (Rhodamine B) is injected into the core of the vortex via a thin gap in the orifice of the vortex cannon. The dashed horizontal line denotes the symmetry axis. (B) Vortex ring radius vs. rescaled time for collisions at various Reynolds numbers. Both cores are dyed, the core centerlines are extracted from 3D reconstructions, and the centerlines are fitted to circles with a fixed center point. The initial time begins when the vortex rings enter the scanning volume and ends when the vortex cores break down. (inset) All experimental curves shifted by $\tilde{t}$ to collapse. The radial growth of the rings coincides with the Biot-Savart prediction.
}
\label{fig:1}
\end{figure}

The geometry of the experimental setup is depicted schematically in Fig.~\ref{fig:1}(A). Two identical counter-rotating vortex rings are fired head-on in a 75-gallon aquarium filled with deionized water, as shown in Movie S1. 
The vortex rings are formed via a piston-cylinder configuration in which a slug fluid with viscosity, $\nu$, is pushed through a cylinder of diameter, D (2.54 cm), at a constant velocity, U, with a stroke, L. The resulting flow is controlled by two dimensionless parameters: the Reynolds number, Re = UD/$\nu$, and the stroke ratio, SR = L/D~\cite{Gharib:1998}.
Fluorescent dye (Rhodamine B) is injected into the cores of the rings as they are formed. Since the collision occurs at a fixed plane in the laboratory frame, this configuration is attractive for directly observing the rapid formation of small-scale flow structures. 
The dynamics and eventual breakdown of the dyed cores are visualized in full 3D by imaging over the collision plane with a scanning laser sheet ($\lambda = 532$~nm), which is pulsed synchronously with a high-speed camera (Phantom V2511). The technical details of how the vortex rings are formed and visualized in 3D are described in previous work~\cite{McKeown:2018} and in supplemental section 1. Additionally, we perform direct numerical simulations (DNS) of interacting vortices at Reynolds numbers equivalent to the experiments (see supplemental sections 1-2 for how the definitions of the Reynolds numbers in simulations and experiments compare).

As the vortex rings collide, they exert mutual strains on one another, causing them to stretch radially at a constant velocity before breaking down at a terminal radius, as shown in Fig.~\ref{fig:1}(B). At low Reynolds numbers, Re $\lesssim$ 5000, the dyed cores break down, ejecting a tiara of secondary vortex rings or smoky turbulent puffs~\cite{Lim:1992,McKeown:2018} at approximately 6 times the initial radius, $R_0$. The initial vortex ring radius and core radius, $\sigma$, were measured separately through particle image velocimetry (PIV), as described in supplemental section 2. Strikingly, for collisions at higher Reynolds numbers, Re $\gtrsim$ 5000, the cores ``burst'' into an amorphous turbulent cloud of dye at a maximum radius of approximately 5$R_0$, indicating the onset of a different breakdown mechanism at this high Reynolds number regime. The mean radial growth of the colliding rings is well described by the Biot-Savart model~\cite{Batchelor:1970}, as described in supplemental section 3(A-B). Additionally, the radial expansion of the rings is encapsulated by a universal functional form, as shown in the inset of Fig.~\ref{fig:1}(B).

While the mean radial growth of the colliding vortex rings follows the same linear evolution at any Reynolds number, the cores themselves develop different forms of perturbations, due to their mutual interaction. 
The formation of these perturbations can arise from two different types of instabilities. The Crow instability~\cite{Crow:1970} causes the cores to develop symmetric circumferential perturbations with long wavelengths, much larger than the core radius, $\sigma$. This instability stems from the mutual advection of the interacting vortices~\cite{Leweke:2016} and governs the breakdown dynamics for collisions at lower Reynolds numbers~\cite{Lim:1992,McKeown:2018, Leweke:2016}. The nonlinear development of the Crow instability causes the rings to deflect into one another and form ``tent-like'' structures~\cite{Hormoz:2012,Brenner:2016,McKeown:2018}, which interact locally at the collision plane. 

\begin{figure}[ht!]
\centering
\includegraphics[width=\linewidth]{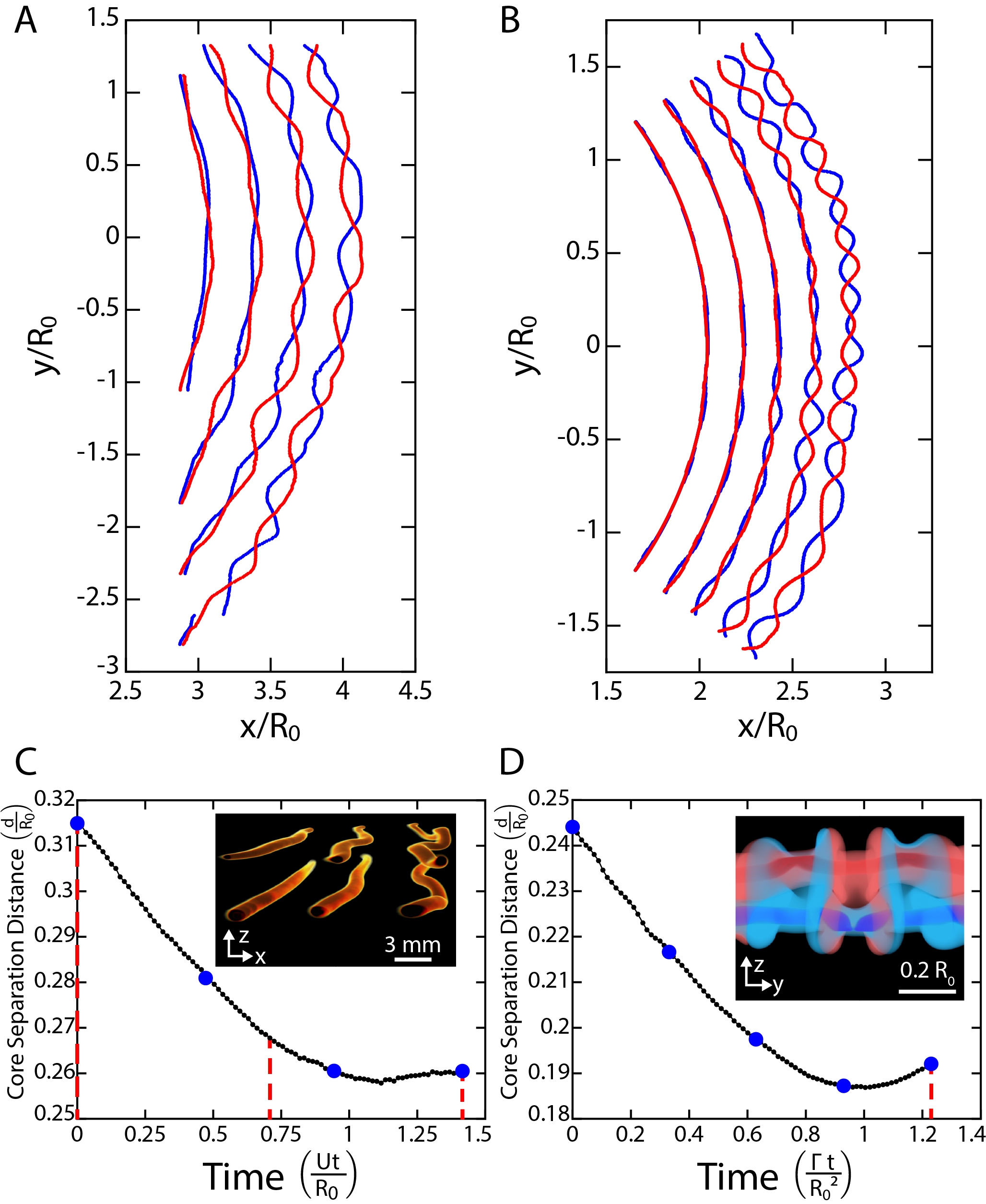}
\caption{
Antisymmetric perturbations in vortex ring collisions. A montage of core centerline trajectories for vortex ring collisions in both (A) experiment and (B) direct numerical simulation. The top ($z>0$) cores are indicated by the red lines, and the bottom ($z<0$) cores are indicated by the blue lines. For the experimental collision, Re = 7000, SR = 2, and $R_0$ = 17.5 mm. For the DNS collision, Re$_\Gamma$ = $\Gamma$/$\nu$ = 4500 and $\sigma = 0.1R_0$.  The cores are segmented from the 3D flow visualization in the experimental collision and from the pressure distribution in the simulation. Mean core separation distance vs. rescaled time for the same (C) experimental and (D) numerical collisions. The blue circles correspond to the trajectories in (A) and (B), and the red dashed lines correspond to the visualizations in the insets. (C, inset) 3D visualization of the dyed vortex cores in the experimental collision. (D, inset) 3D visualization of the dyed vortex rings in the simulation, showing both the dye in the cores (dark) and surrounding them (light).  
}
\label{fig:2}
\end{figure}

At higher Reynolds numbers, both our experiments and simulations show that the breakdown dynamics are governed by the elliptical instability, causing the vortex cores to develop short-wavelength perturbations on the order of the core radius~\cite{Tsai:1976,Moore:1975, Schaeffer:2010} (see supplemental section 3(C)). This instability originates from the parametric excitation of Kelvin modes in the vortex cores due to the resonant interaction of the strain field from the other vortex~\cite{Leweke:2016, Kerswell:2002}. A hallmark of the elliptical instability, these short-wavelength perturbations grow synchronously in an antisymmetric manner, as shown for two typical  experimental and numerical examples in Fig.~\ref{fig:2}(A-B). 

As the elliptical instability grows radially along the collision plane, the mean spacing between the cores, $d$, decreases linearly. However, the separation distance between the cores saturates when the perturbations deflect out-of-plane, just prior to breaking down, as shown in Fig.~\ref{fig:2}(C-D) and Movie S2.
The minimum mean spacing between the cores is approximately equal to twice the initial core radius, $ \sigma$. For the experiment, $\sigma = 0.14 \pm 0.01 ~R_0$, and for the simulation, $\sigma  = 0.1 R_0$ (see supplemental section 2).
After the elliptical instability develops and the symmetry of the two cores is broken, a periodic array of satellite flow structures is shed from each core, bridging the gap between them, as shown in the inset of Fig.~\ref{fig:2}(D) and Movie S3. Notably, in our experiments, the emergence of these secondary flow structures can only be resolved if the fluid outside of the cores is dyed.

\begin{figure*}[ht]
\centering
\includegraphics[width=0.8\linewidth]{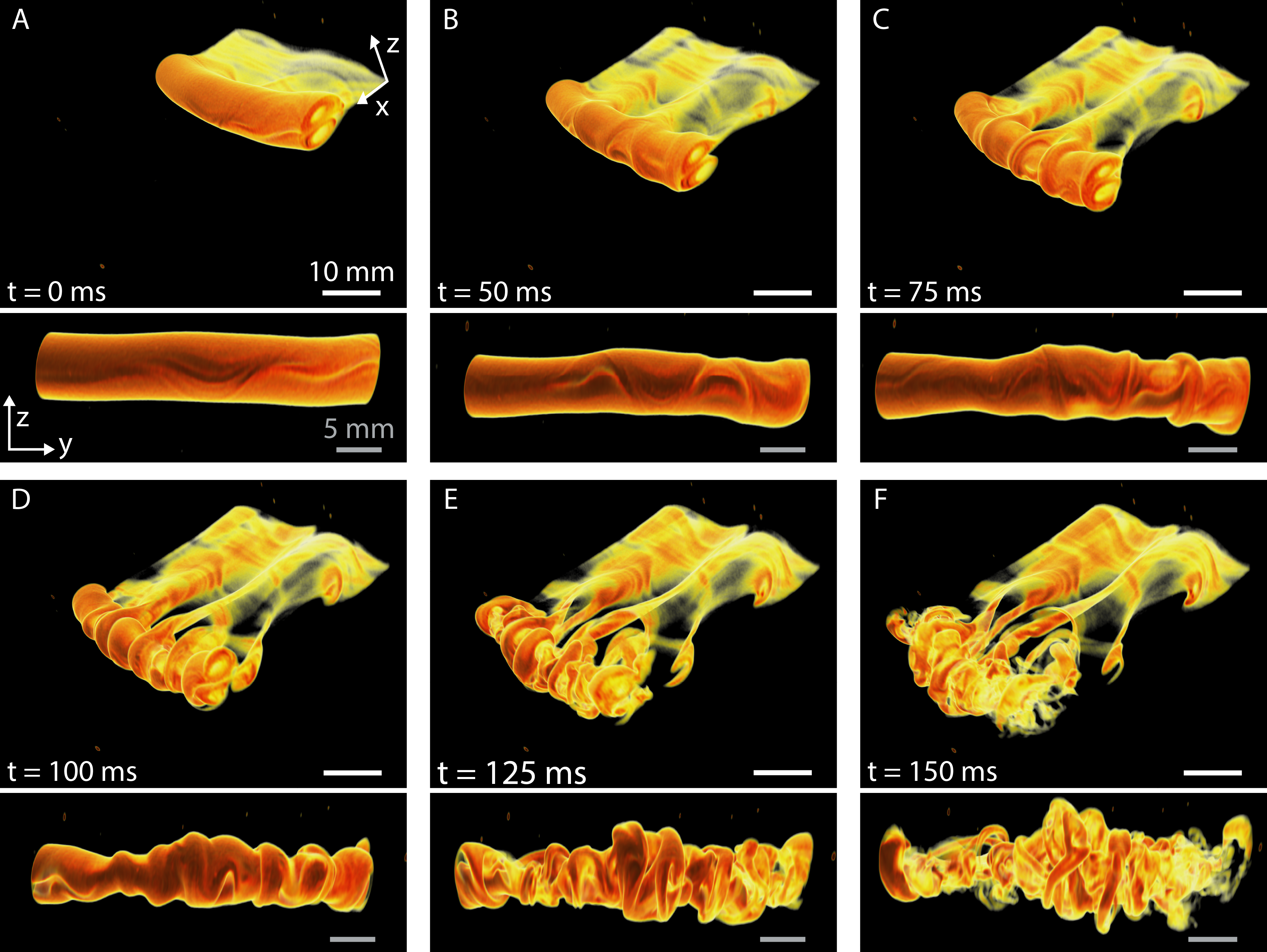}
\caption{
Formation of perpendicular secondary filaments in a typical experimental vortex ring collision. 3D reconstruction of two fully-dyed vortex rings colliding head-on, viewed from overhead (top) and from the side (bottom). Re = $6000$ and SR = $2.5$. (A-C) As the rings grow, they interdigitate as the dye from the upper ring is wrapped around the lower ring and vice-versa. (D-E) The colliding rings form an array of secondary vortex filaments that are perpendicular to the vortex cores. (F) The cores and perpendicular filaments break down into a fine-scale turbulent cloud.}
\label{fig:3}
\end{figure*}

In order to better resolve the late-stage development of the elliptical instability and the resulting breakdown, we dye the full vortex rings, as shown in Fig.~\ref{fig:3}(A-F) and Movie S4. 
Observing the fully dyed vortex rings reveals the intricate structure of the flow that develops in response to the core dynamics.
The antisymmetric coupling of the perturbations break the azimuthal symmetry of the flow, leading to the exchange of fluid between the two rings. 
This periodic wrapping of dye causes the outer layers of the rings to interdigitate around one another along alternating ``tongues,''~\cite{Leweke:1998} as shown in Fig.~\ref{fig:3}(A-B). 
At the boundaries of adjacent tongues, the dyed fluid curls into vortex filaments, perpendicular to the cores, as shown in Fig.~\ref{fig:3}(C).
These alternating filaments are stretched by the circulating vortex cores into an array of counter-rotating secondary vortices, as shown in Fig.~\ref{fig:3}(D-E). The secondary filaments have a fleeting lifetime of only tens of milliseconds before they break down. 
Violent interactions between the secondary filaments and primary cores result in the rapid ejection of fine-scale vortices and the formation of a turbulent cloud, as shown in Fig.~\ref{fig:3}(F).

\begin{figure}
\centering
\includegraphics[width=\linewidth]{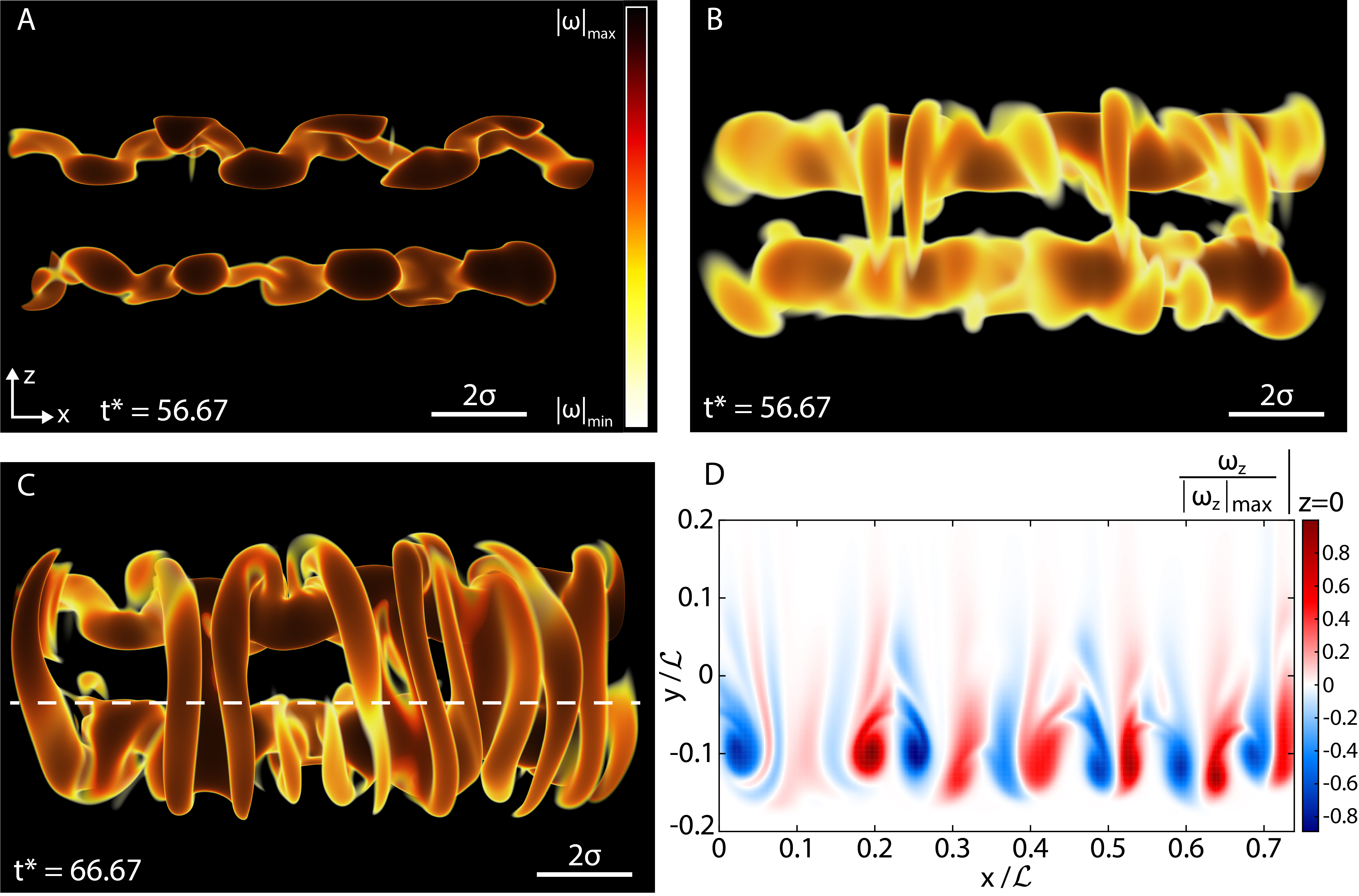}
\caption{
Generation of perpendicular secondary filaments in DNS. (A-C) Vorticity modulus for simulated interacting tubes where Re$_\Gamma=4500$, $\sigma = 0.06\mathcal{L}$, $b = 2.5\sigma$, and $t^{*} = \Gamma t / b^2$. The vorticity modulus is normalized by the maximum vorticity modulus during the simulation, $|\omega|_{\text{max}}$. (A) The initial antisymmetric perturbations of the cores develop as the tips of the perturbations locally flatten ($0.103 \leq |\omega|/|\omega|_{\text{max}} \leq 0.117$). (B) At the same time, low-vorticity perpendicular filaments form as a result of the perturbations ($0.046 \leq |\omega|/|\omega|_{\text{max}} \leq 0.092$). (C) Once the secondary filaments form, their vorticity amplifies ($0.076 \leq |\omega|/|\omega|_{\text{max}} \leq 0.114$). (D) Vorticity distribution in the $z$-direction along the center plane ($z=0$) indicated by the dashed line in (C). Adjacent secondary filaments counter-rotate. 
}
\label{fig:4}
\end{figure}

By performing direct numerical simulations of the colliding vortices, we additionally probe how energy is transferred through the flow via the onset of the elliptical instability.
Since the breakdown of the vortices is localized to the area around the cores, we implement a new configuration for the simulations, which consists of two initially parallel, counter-rotating vortex tubes with circulation $\Gamma$, initially spaced a distance, $b = 2.5\sigma$, apart. The flow is simulated in a cubic domain of side length, $\mathcal{L}=16.67\sigma$, and the Reynolds number of this configuration is given by Re$_\Gamma = \Gamma/\nu$ (see supplemental section 1). From PIV measurements, we find that Re$_\Gamma \approx 0.678$Re as shown in supplemental section 2. 
The dynamics of the vorticity distribution in the simulated flow are qualitatively equivalent to the experimental flow visualizations, as shown for a typical example at Re$_\Gamma = 4500$ in  Fig.~\ref{fig:4}(A-C) and Movie S5.
When the antisymmetric perturbations resulting from the elliptical instability materialize, the tips of the perturbed cores deform into flattened vortex sheets, as illustrated in Fig.~\ref{fig:4}(A). These sheets are stretched by strains applied by the other core and roll up along the edges into an alternating series of hairpin vortices, as shown in Fig.~\ref{fig:4}(B) and supplemental section 4(A).
Upon stretching across the gap to the other perturbed core, these hairpin vortices form an ordered array of secondary vortex filaments perpendicular to the initial tubes, as shown in Fig.~\ref{fig:4}(C). Adjacent pairs of secondary filaments counter-rotate relative to one another~\cite{Leweke:1998}, as shown in Fig.~\ref{fig:4}(D). 
Integrating the transverse vorticity along the symmetry plane, we find that as the secondary filaments are stretched, approximately $25\%$ of the streamwise circulation from the initial vortex tubes is conveyed to each filament (see supplemental section 4(B)). 
As the vorticity of the original tubes is transferred to the secondary filaments, the circulation of the flow is conserved.

\begin{figure*}
\centering
\includegraphics[width=0.95\linewidth]{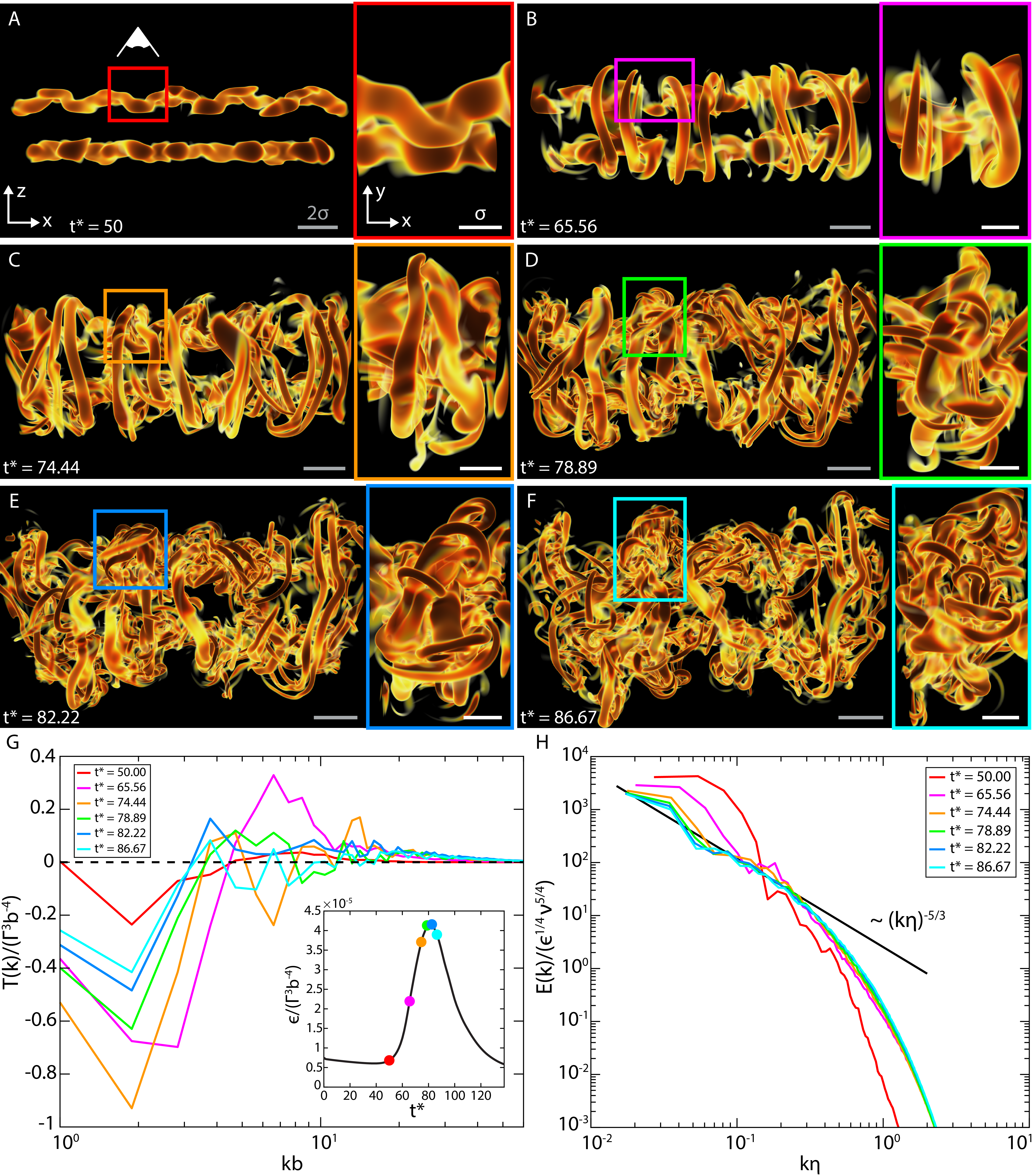}
\caption{
The development of a turbulent cascade. (A-F) Vorticity modulus for simulated interacting tubes where Re$_\Gamma = 6000$, $\sigma = 0.06\mathcal{L}$, $b = 2.5\sigma$, and $t^{*} = \Gamma t / b^2$. Each panel shows the front view of the full cores (left) and a close-up top view of the interacting secondary filaments indicated in the full view (right).  (A) The antisymmetric perturbations of the cores develop. (B) Perpendicular secondary filaments form between the cores. (C) Secondary filaments begin to interact with each other and break down. (D) Tertiary filaments begin to form perpendicular to the secondary filaments. (E) Tertiary filaments are fully formed. (F) The flow breaks down into a disordered tangle of vortices. The vorticity thresholds are $0.079 \leq |\omega|/|\omega|_{\text{max}} \leq 0.099$ for (A) and $0.110 \leq |\omega|/|\omega|_{\text{max}} \leq 0.211$ for (B-F), where $|\omega|_{\text{max}}$ is the maximum vorticity modulus for the entire simulation. (G) Normalized shell-to-shell energy transfer spectra indicate whether a mode is an energy source ($T(k)>0$) or sink ($T(k)<0$). At early times ($t^{*} = 65.56, 74.44$), the secondary filaments switch from energy sinks to sources as they are generated and then interact to form new vortices. At late times, the spectra flatten as energy is transferred more uniformly across the scales of the flow. (G, inset) Normalized kinetic energy dissipation rate as a function of time.  The energy dissipation rate increases with the development of the secondary filaments and peaks as the secondary filaments and residual cores break down into a tangle of fine-scale vortices. (H) Normalized kinetic energy spectra show the rapid development of a sustained turbulent state with Kolmogorov scaling--as indicated by the black line--around the peak dissipation rate.
}
\label{fig:5}
\end{figure*}

Once formed, each pair of secondary filaments can be locally viewed as a replica of the initial flow configuration on a smaller scale and with a reduced circulation, hence corresponding to a smaller effective Reynolds number. 
The resulting close-range interactions of neighboring filaments can lead to an iterative cascade by which even more generations of small-scale vortices are formed. 
For collisions at moderately high Reynolds numbers (e.g. Re$_\Gamma =3500$), the concentrated strains exerted by the counter-rotating secondary filaments cause one of them to flatten into an extremely thin vortex sheet and split into two smaller tertiary vortex filaments, as shown in Movies S6-S7 and supplemental section 5. This behavior is consistent with the breakdown mechanism observed experimentally in the head-on collision of vortex rings at comparatively lower Reynolds numbers mediated by the Crow instability~\cite{McKeown:2018}.

In the high-Reynolds number limit, the secondary filaments may give rise to another generation of perpendicular tertiary vortex filaments, as shown for a typical example at Re$_\Gamma =6000$ in Fig.~\ref{fig:5}(A-F) and Movies S8-S9. 
The secondary filaments are drawn into one another due to their mutual counter-rotation, as shown in the close-ups in Fig.~\ref{fig:5}(B-C). 
The narrow gap between these writhing vortices, which experiences intense strain, almost instantaneously becomes enveloped by several high-vorticity tertiary filaments, as shown in Fig.~\ref{fig:5}(D-E). 
These tertiary filaments align perpendicular to the previous generation of vortices, wrapping tightly around them. The tertiary filaments develop locally in an ordered manner while the remnants of the primary cores and secondary filaments become increasingly entangled into a disordered ``soup'' of vortices, as shown in Fig.~\ref{fig:5}(F).

The iterative breakdown process occurs over diminutive length scales and fleeting time scales. This rapid generation of small-scale vortices leads to a dramatic increase in the energy dissipation rate, $\epsilon$, as shown in the inset of Fig.~\ref{fig:5}(G).
The initial increase in $\epsilon$ is triggered by the onset of the elliptical instability and the formation of antisymmetric perturbations in the cores. The following precipitous rise in the dissipation rate coincides with the formation of the perpendicular secondary filaments. 

The rate at which kinetic energy is transferred across scales is calculated for the simulation in three-dimensional Fourier space through the instantaneous shell-to-shell energy transfer spectrum, $T(k,t)$, as shown in Fig.~\ref{fig:5}(G) ~\cite{Pope:2000,Lin:1947}. At a fixed time, $T(k)$ 
is positive for a wavenumber, $k$, when energy flows toward the corresponding spatial scale $(\sim k^{-1})$. Conversely, a negative value of $T(k)$ indicates the flow of energy away from that corresponding spatial scale to other modes (see supplemental section 6 for details). 
When initially formed, the secondary filaments become pronounced energy sinks, given the large positive value of $T(k)$ at the intermediate wavenumber of approximately $kb = 6.75$. This coincides with their absorption of energy from the primary vortex cores.
Next, as the secondary filaments become fully developed and interact with each other, they change behavior and become sources of energy, as indicated by the negative value of $T(k)$. Coupled with the simultaneous increase in the dissipation rate, this change in behavior of the secondary filaments from energy sinks to energy sources indicates the existence of a cascade by which kinetic energy is conveyed to smaller scales. 

Through the breakdown of the secondary filaments and residual vortex cores into a disordered tangle of vortices, the dissipation rate reaches a maximum value. 
At this point, the flow is most vigorous and the energy transfer spectra asymptote toward a flattened profile, indicating that energy is conveyed more uniformly across the various scales of the system, as shown in Fig.~\ref{fig:5}(G).
Thus, for this brief time, $\epsilon$ maintains an approximately constant maximum value, as the energy is smoothly transferred to the smallest, dissipative scale, $\eta = \epsilon^{-\frac{1}{4}} \nu^{\frac{3}{4}}$. Kolmogorov proposed that for turbulent flows under similar conditions, the kinetic energy spectrum follows a distinct scaling of $(k \eta)^{-5/3}$~\cite{Kolmogorov:1941}.
Strikingly, we find that the fully developed turbulent cloud formed by the collision of the two vortices, indeed, exhibits Kolmogorov scaling. 
The evolution of the normalized energy spectra, $E(k)/(\eta^{\frac{1}{4}} \nu^{\frac{5}{4}})$, demonstrates how the flow reaches a sustained turbulent state around the peak dissipation rate, as shown in Fig.~\ref{fig:5}(H). 
This turbulent energy spectrum scaling at the peak dissipation rate also emerges during the breakdown of interacting vortex tubes mediated by the elliptical instability at lower Reynolds numbers, as shown in supplemental section 7. 
Since the energy input of the system is finite, this turbulent state cannot be maintained indefinitely.
As time progresses further, the viscosity of the fluid damps out the motion of the vortices at the smallest scales. 
While much energy remains at the large scales of the flow, it is unable to be transmitted to smaller scales following this iterative breakdown.
Accordingly, the energy dissipation rate decreases and the turbulent state decays.

The violent interaction between two counter-rotating vortices leads to the rapid emergence of a turbulent cascade, resulting in a flow with an energy spectrum that--for an ethereal moment--obeys Kolmogorov scaling. We find that the emergence of this turbulent cascade is initiated by the late-stage, nonlinear development of the elliptical instability, which forms an ordered array of counter-rotating secondary vortex filaments perpendicular to the primary cores. 
In the high-Reynolds number limit, the neighboring secondary filaments may interact to form a new generation of perpendicular tertiary vortex filaments. 
These interactions of the secondary filaments with each other and the remnants of the vortex cores lead to the rapid formation of small-scale vortices. This ensemble of vortices interacting over the full range of scales of the system provides a conduit through which energy cascades down to the dissipative scale. 

The iterative cascade, which leads to the generation of vortices at decreasingly small length scales, is strongly reminiscent of the mechanism proposed by Brenner, Hormoz, and Pumir~\cite{Brenner:2016}. One may speculate that the self-similar process suggested by this work could be modeled by assuming, in the spirit of~\cite{Brenner:2016}, that at each iteration, the circulation is multiplied by a factor $x_\Gamma < 1$, and the characteristic scale of the vortices by a factor $x_\delta < 1$, resulting after $n$ steps in a generation of vortices with circulation, $\Gamma_n = x_\Gamma^{n-1} \Gamma_1$, and a spatial scale, $\delta_n = x_{\delta}^{n-1} {\delta}_1$. 
The corresponding time scale over which each step evolves can be estimated as $ t_n \sim \delta_n^2/\Gamma_n \sim (\delta_1^2/\Gamma_1) \, (x_{\delta}^2/x_\Gamma)^{n-1}$. The cascade can go all the way down to vanishingly small spatial scales in a finite time provided $x_{\delta}^2 < x_\Gamma$. The numerical results presented here, during the first steps of the cascade, suggest that $x_\Gamma \sim 0.25$, and $x_{\delta} \sim 0.2-0.4$, and therefore that the cascade may proceed in a finite time. It would be interesting to understand whether the cascade suggested by this work proceeds faster or slower than the Kolmogorov cascade. Whereas Kolmogorov theory implies $t_n/t_1 \sim (\delta_n/\delta_1)^{2/3}$~\cite{Frisch:1995}, our results imply that $t_n/t_1 \sim (\delta_n/\delta_1)^{2 - \ln(x_\Gamma)/\ln(x_\delta)}$. Therefore, the cascade proposed in this work proceeds faster than the Kolmogorov cascade for $x_\delta < x_\Gamma^{3/4}$. Our estimates suggest that the two cascades may proceed asymptotically at a comparable rate. A more precise understanding of the development of the elliptical instability is necessary to determine accurately the scaling factors $x_\Gamma$ and $x_\delta$. 

The essential element of the cascade process is that at each scale, discrete pairs of antiparallel vortices are able to locally interact and produce a subsequent iteration via the elliptical instability. 
Vortices of similar size and circulation locally align in an antiparallel manner when they interact. This is a well established consequence of Biot-Savart dynamics~\cite{Siggia:1985}. 
Thus, the largest strains that drive the cascade will arise from the interactions of nearby vortices. we suggest that iterations of this cascade could proceed down to ever-smaller scales until viscous effects take over. 
Yet, the proliferation of other small-scale vortices, clearly visible in Fig. 5, could conceivably prevent vortex pairs from forming at some stage of the process, and we do not rule out that it may influence the dynamics. 
We remark, however, that the present work shows that only two clear iterations of the cascade are sufficient to produce a Kolmogorov spectrum. 
Even in the high-Reynolds number limit, a finite set of iterations occurring simultaneously for many independent pairs of interacting vortices might suffice to establish and sustain a turbulent cascade. The details of how this iterative process unfolds in the limit of large Reynolds number is an important question for future research.

This framework strongly agrees with recent works by Goto et al. in a fully turbulent flow regime. Namely, their numerical results demonstrate the existence of many independent pairs of antiparallel vortices interacting and locally forming smaller generations of perpendicular vortex filaments in both fully developed homogeneous isotropic turbulence and wall-bounded turbulence~\cite{Goto:2012,Goto:2017,Motoori:2019}.
These discrete interactions of antiparallel vortex pairs appear simultaneously throughout Goto's simulations over four distinct scales~\cite{Goto:2017}.
Due to the striking similarities between the iterative mechanism we observe and the results of Goto, we propose that the elliptical instability is likely the means by which these successive generations of perpendicular filaments are formed.
Establishing a precise connection between our own results and Goto's observations requires a fully quantitative analysis, which is beyond the scope of the present work.

Our work thus demonstrates how the elliptical instability provides a long-sought-after mechanism for the formation and perpetuation of the turbulent energy cascade through the local interactions of vortices over a hierarchy of scales. Supplied by the injection of energy at large scales, discrete iterations of this instability effectively channel the energy of a flow down to the dissipative scale through the formation of new vortices. 
From a quantitative point of view, the approximate estimates provided in this work suggest that the corresponding cascade proceeds in a finite time, although a precise comparison with the Kolmogorov cascade requires a better understanding of the nonlinear development of the elliptical instability.
While the dynamics of turbulent flows likely involve other multi-scale vortex interactions, this fundamental mechanistic framework can begin to unravel the complexity that has long obscured our understanding of turbulence. 

\section{Methods}
Technical details for the experiments and simulations are provided in supplemental section 1.

\section{Acknowledgements}
This research was funded by the National Science
Foundation through the Harvard Materials Research Science and Engineering Center DMR-1420570 and through the
Division of Mathematical Sciences DMS-1411694 and
DMS-1715477. M.P.B. is an investigator of the Simons
Foundation. R.O.M. thanks the Core facility for Advanced Computing and Data Science (CACDS) at the University of Houston for providing computing resources.


\bibliography{ms}
\clearpage

\end{document}


\title{Turbulence generation through an iterative cascade of the elliptical instability \\ Supplementary Information}

\author{Ryan McKeown$^1$, Rodolfo Ostilla-M\'{o}nico$^{2,*}$, Alain Pumir$^3$, Michael P. Brenner$^{1,4}$, and  Shmuel M. Rubinstein$^{1,*}$}

\affiliation{$^1$ School of Engineering and Applied Sciences, Harvard University, Cambridge, MA 02138, USA \\
$^2$ Department of Mechanical Engineering, University of Houston, Houston, TX 77204, USA \\
$^3$ Univ Lyon, ENS de Lyon, Laboratoire de Physique, F-69342 Lyon, France \\
$^4$ Google Research, Mountain View, CA 94043, USA \\
$^*$ Corresponding Authors: rostilla@central.uh.edu, shmuel@seas.harvard.edu }

\maketitle

\tableofcontents

\newpage


\section{Methods and materials}
\label{sec:Methods}

We use both experiments and simulations to probe the dynamic formation of the turbulent cascade resulting from the interaction between counter-rotating vortices. Experimentally, we examine the head-on collision of vortex rings, and numerically we examine the collision of vortex rings and vortex tubes. 
In the experiments~\cite{McKeown:2018}, fluorescent dye (Rhodamine B) is injected into the initially formed rings to visualize the core dynamics, as shown in Fig. 1(A) in the main text. Two vortex rings are fired head-on into one another in a 75-gallon aquarium. The vortex cannons are positioned a distance of $8D$ apart, where $D$ is the vortex cannon diameter. The full, three-dimensional dynamics of the resulting collision are visualized by tomographically scanning over the collision plane with a rapidly translating pulsed laser sheet ($\lambda = 532$ nm). The pulsing of the  laser (Spectraphysics Explorer One 532-2W) is synchronized with the exposure signal of a high speed camera (Phantom V2511), which images the illuminated plane head-on. Each image plane spans along the $xy$-plane, and the laser sheet scans in the $z$-direction. Thus, for each scan, the image slices are stacked together to form a 3D reconstruction of the collision. The spatial resolution of each volume is 145 $\times$ 145 $\times$ 100 ~$\mu$m$^3$ per voxel in ($x$,$y$,$z$), and the time resolution is up to 0.5 ms per scan. The number of voxels in each scanned volume depends on the imaging window size, the camera frame rate, and the scanning rate. For example, the volume size is 512 $\times$ 512 $\times$ 64 voxels in $(x,y,z)$ for the dyed core collision in Movie~\ref{mov:exp_cores} and 384 $\times$ 288 $\times$ 75 voxels for the fully dyed ring collision in Movie~\ref{mov:exp_rings_dye}.  
The series of volumetric scans are reconstructed in full 3D with temporal evolution using Dragonfly visualization software (Object Research Systems).
The imaging apparatus can only detect the dyed regions of the fluid, so any flow structures that emerge during the breakdown that are undyed cannot be observed. When only the vortex cores are dyed, we probe through the volumes of each 3D scan along the azimuthal direction to locate the centroids of the cores at each cross section. This enables us to extract the vortex core centerlines, which we use to track the deformation of the cores and measure the vortex ring radius, $R(t)$, and the average spacing between the cores, $d(t)$. In order to visualize the development of secondary flow structures that emerge during the collisions, we fill the vortex cannons with fluorescent dye prior to driving the pistons to form the rings. As a result, the regions around the vortex cores are dyed, as shown in Fig. 3 in the main text.

We use  direct numerical simulations (DNS) to further probe the unstable interactions between the vortices. This allows us to directly examine the evolution of the vorticity field, relate it to experimental visualizations, and compute statistical quantities characterizing the flow. We solve the incompressible Navier-Stokes equations using an energy-conserving, second-order centered finite difference scheme with fractional time-stepping. We implement a third-order Runge-Kutta scheme for the non-linear terms and second-order Adams-Bashworth scheme for the viscous terms  \cite{Kim:1985,Verzicco:1996}. We simulate both vortex rings and tubes, in cylindrical and Cartesian coordinates, respectively. To avoid singularities near the axis, the cylindrical solver uses $q_r = r v_r$ as a primitive variable~\cite{Verzicco:1996}. The time-step is dynamically chosen such that the maximum Courant-Friedrich-Lewy (CFL) condition number is 1.2. Resolution adequacy is checked by three methods: monitoring the viscous dissipation and the energy balance, examining the Fourier energy spectra, and using the instantaneous Kolmogorov scale. White noise is added to all all initial conditions to trigger the most unstable modes. 

The rings are initialized as two counter-rotating Gaussian (Lamb-Oseen) vortices, each with a core radius $\sigma$ wrapped into a torus of radius $R_0$. The control parameters for this system are the circulation Reynolds number, Re$_\Gamma = \Gamma/\nu$ and the slenderness ratio of the rings, $\Lambda = \sigma/R_0$. The circulation of the vortex rings, $\Gamma$, and the initial ring radius, $R_0$, are used to non-dimensionalize parameters in the code. We simulate the collision in a closed cylindrical domain, bounded by stress-free walls at a distance far enough to not affect the collision. After testing several configurations, the bounds on the domain were placed a distance $R_0$ below and above the rings, and $5R_0$ from the ring axis in the radial direction. For the simulation presented in this paper, we selected a ring slenderness of $\Lambda=0.1$, a circulation Reynolds number of Re$_\Gamma=4500$, and an initial ring-to-ring distance of $d=2.5R_0$. These parameters are comparable to the experimental vortex rings, as shown by the measurements in supplemental section~\ref{sec:PIV}. Points are clustered near the collision regions in the axial and radial directions, while uniform resolution is used in the azimuthal direction \cite{Gharib:1998,McKeown:2018}. A rotational symmetry of order five is forced on the simulation to reduce computational costs. The vortex core centerlines are located by slicing azimuthally through the pressure field at every time step and identifying the local minima of each vortex cross section. Additionally, a simulated passive scalar is injected into the vortex rings to visualize the dynamics of the collision and compare with experiments, as shown in Movie~\ref{mov:sim_rings_dye}. Due to computational restrictions, the diffusivity of the dye is equal to the kinematic viscosity of the fluid (i.e. the Schmidt number is unity).

For the vortex tubes, we simulate a triply periodic cubic domain of period $\mathcal{L}$, which is discretized using a uniform grid. The two counter-rotating, parallel tubes are both initialized with a Gaussian (Lamb-Oseen) vorticity profile of radius, $\sigma$, and circulation, $\Gamma$, initially separated a distance, $b$, apart. The system is characterized by two dimensionless parameters: the circulation Reynolds number, Re$_\Gamma$, and the ratio $\sigma/b$. Again, the circulation, $\Gamma$, is used as a non-dimensional parameter, along with $b$. We set the core size to $\sigma = 0.06 \mathcal{L}$, fix $b/\sigma = 2.5$, and run simulations with Re$_\Gamma$ at $2000$, $3500$, $4500$, and $6000$, with grid sizes of $256^3$, $360^3$, $540^3$, and $540^3$, respectively. As the counter-rotating tubes interact and break down, they naturally propagate through the periodic domain. For each visualization, the propagation of the tubes is subtracted so that the tubes remain in the center of the domain. Additionally, in all 3D visualizations of the vorticity modulus, $|\omega|(t)$, the vorticity modulus at each voxel is normalized by the maximum vorticity modulus for all time, $|\omega|(t)_{\text{max}}$ (see supplemental section~\ref{subsec:Vorticity_Across_Re}).

\section{PIV analysis of vortex ring geometry}
\label{sec:PIV}
The  vortex rings are characterized experimentally through 2D patricle image velocimetry (PIV). The fluid is seeded with polyamide particles with a diameter of $50 ~\mu$m and a density of 1.03 g/mL (Dantec Dynamics). 
A laser sheet is positioned along the central axis of the vortex cannon in order to illuminate the cross section of the ejected vortex rings. The motion of the particles in the vortex rings along this cross section is imaged with a high-speed camera (Phantom V2511) with a window size of 1280 $\times$ 800 pixels at a maximum frame rate of 2000 fps, such that the resolution is 0.12 mm/pixel. 
Vortex rings are formed over a range of stroke ratios (SR = $L/D$) and Reynolds numbers (Re$ = UD/\nu$), where $L$ is the stroke of the piston, $D= 25.4$ mm is the diameter of the vortex cannon, $U$ is the piston velocity, and $\nu$ is the kinematic viscosity of water.

\begin{figure}[ht!]
\centering
\includegraphics[width=\textwidth]{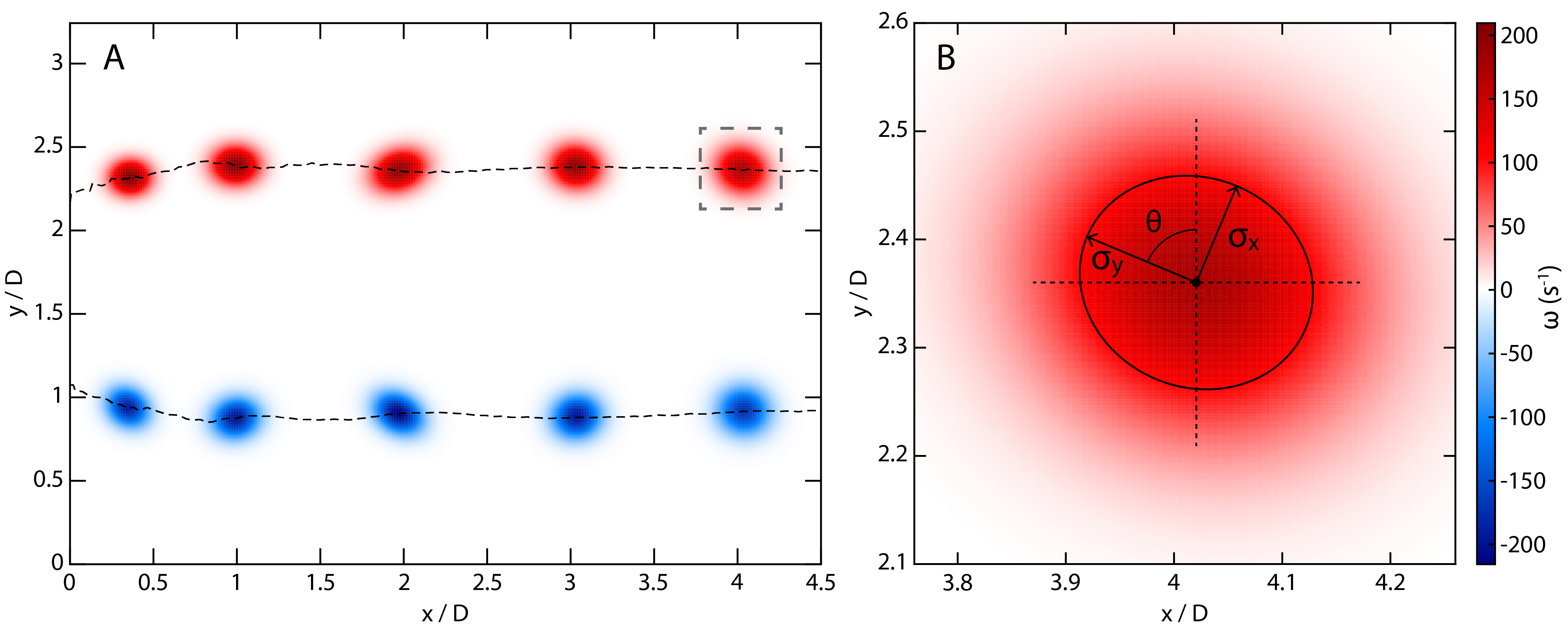}
\caption{
Vortex ring tracking and measurement through PIV. (A) 2D vorticity distribution and core trajectories for a vortex ring formed 
via a piston-cylinder assembly, where  Re = $7000$ and SR = $2.5$. Each vorticity distribution results from fitting the raw vorticity data to a 2D Gaussian function. The time steps correspond to $t = 0, 0.158, 0.368, 0.578, ~$and$~ 0.788$ seconds, and the vortex cannon orifice is located at $x=0$. (B) Zoomed-in view of the final vortex core indicated by the dashed gray box. The vortex core is defined by the level curve shown by the black ellipse, with a center point, $(\mu_x, \mu_y)$, a rotation angle, $\theta$, and standard deviations, $\sigma_x$ and $\sigma_y$.
}
\label{fig:PIV}
\end{figure}

The velocity field for each generated vortex ring is calculated using MATLAB PIVsuite. The cores of the vortex rings are identified by computing the vorticity field, and each core is fitted to a two-dimensional Gaussian function, as shown for a typical example in Fig.~\ref{fig:PIV} and Movie~\ref{mov:PIV}.
After pinching off, the vortex ring reaches a steady size with radius, $R_0$, as it propagates forward through the fluid, as shown in Fig.~\ref{fig:PIV}(A). Additionally, the core radius, $\sigma$, is calculated by averaging the standard deviations of each Gaussian fit, $\sigma_x$ and $\sigma_y$, as shown in Fig.~\ref{fig:PIV}(B).

\begin{figure}[t]
\centering
\includegraphics[width=\textwidth]{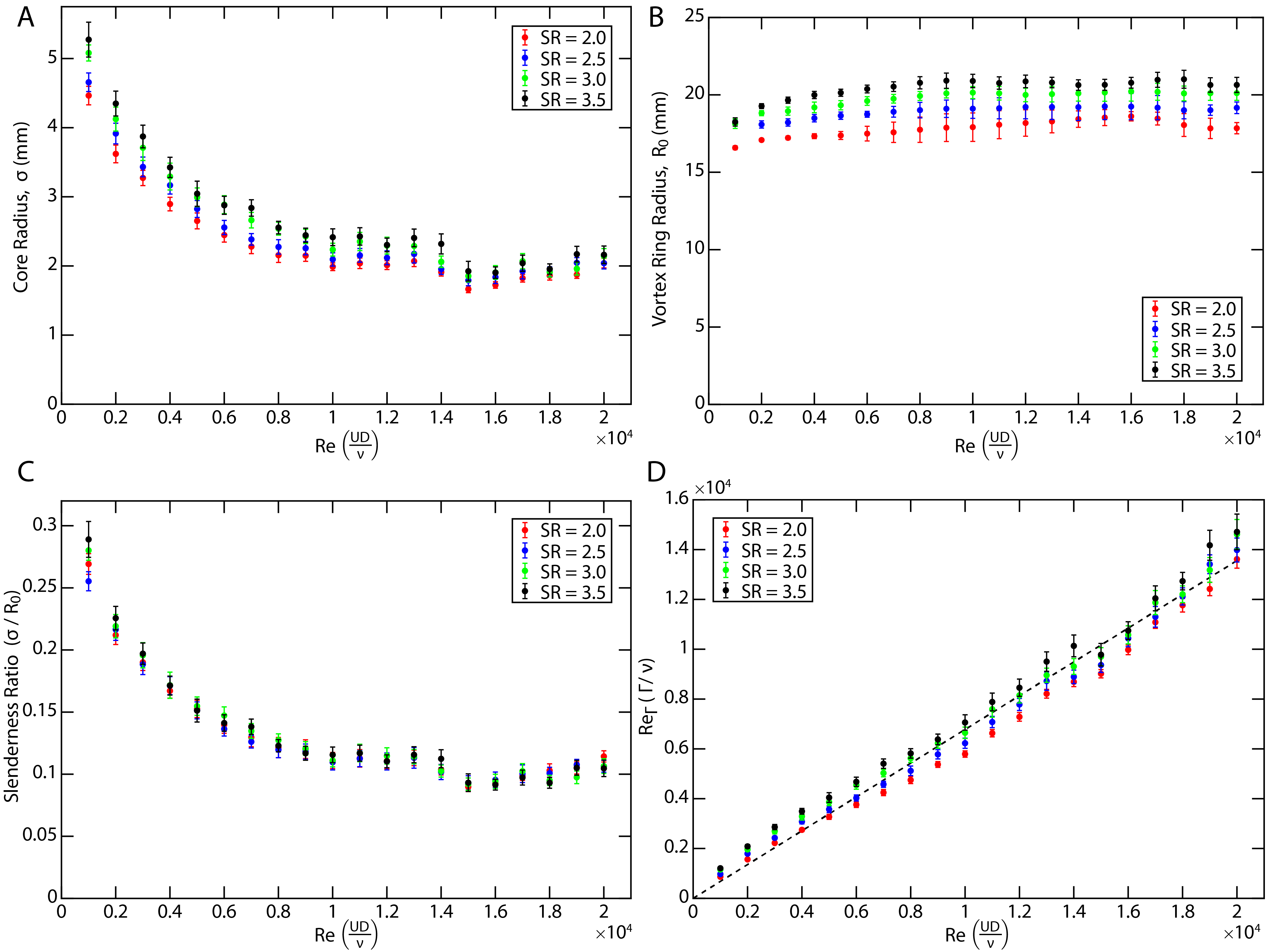}
\caption{
 Vortex ring and core geometry. From the Gaussian fits of the vorticity data, the following parameters are measured as a function of the Reynolds number across various stroke ratios: (A) vortex core radius, $\sigma$, (B) vortex ring radius, $R_0$, (C) slenderness ratio, $\Lambda$, and (D) circulation Reynolds number, Re$_\Gamma$. The dashed line in (D) corresponds to the linear fit to the data: Re$_{\Gamma} = 0.678$Re. All calculations are performed when the vortex rings reach a distance between $2D$ and $4D$ from the vortex cannon orifice.
}
\label{fig:Vortex_Data}
\end{figure}

By fitting to the vortex cores to a Gaussian function, we evaluate the geometry of the rings over a wide range of Reynolds numbers at various stroke ratios, as shown in Fig.~\ref{fig:Vortex_Data}. 
This parameterization of the vortex rings and their core structures enables us to directly relate the initial state of the vortex rings used in the experimental collisions to those of the simulations, as described in the main text. 
Accordingly, all of the parameterizations for the vortex ring geometry are performed when the vortex ring propagates a distance between $2D$ and $4D$ from the orifice of the vortex cannon. This corresponds to the range of distances required for the vortex ring to pinch off and reach a steady morphology prior to reaching where the collision plane is located in the experiments at $4D$. 

At every stroke ratio, the core size is larger for the vortex rings produced with lower Reynolds numbers, as shown in Fig.~\ref{fig:Vortex_Data}(A). However, for vortex rings with a Reynolds number greater than $\sim 10,000$, the core size remains relatively constant. 
This is because vortex rings formed at lower Reynolds numbers are more susceptible to viscous dissipation from the ambient fluid as they propagate forward. 
This dissipation leads to a spreading of the vorticity distribution in the cores through the diffusion of momentum via the viscosity of the fluid, thereby resulting in the thicker cores of the vortex rings at lower Reynolds numbers. 
The vortex ring radius, $R_0$, remains roughly constant for each stroke ratio over this wide range of Reynolds numbers, as shown in Fig.~\ref{fig:Vortex_Data}(B). Like with the core size, the vortex rings formed with a larger stroke ratio naturally have a slightly larger radius, as more fluid is injected to form vortex rings. 
However, by normalizing the core radius with the vortex ring radius at each Reynolds number to compute the slenderness ratio, $\Lambda$, the data collapses, as shown in Fig.~\ref{fig:Vortex_Data}(C). 
This consistency of the vortex ring and core geometry across various stroke ratios informes the selection of the vortex ring parameters in the simulated collisions described in the main text. 
In particular, for all of the simulated vortex ring collisions, a slenderness ratio of $\Lambda = 0.1$ is used.

The circulation of the vortex cores, $\Gamma$, is calculated by integrating the raw vorticity data for each core over an elliptical contour defined by the Gaussian fit that overcompensates the core size by a factor of 1.5 for both standard deviations. 
The magnitude of the circulation is then normalized by the kinematic viscosity of the fluid in order to compute the circulation Reynolds number, Re$_{\Gamma} = \Gamma/\nu$. This parameter, as previously discussed in the main text and in supplemental section~\ref{sec:Methods}, is used to define the Reynolds number in the numerical simulations of the colliding vortex rings and the interacting vortex tubes. 
This calculation thus allows to relate the Reynolds number of the vortex rings formed experimentally with those formed numerically, as shown in Fig.~\ref{fig:Vortex_Data}(D). 
Accordingly, we find that the experimental Reynolds number scales linearly with the circulation Reynolds number, such that Re$_{\Gamma} \approx 0.678$Re, which is consistent with previous experimental works ~\cite{Gharib:1998}.

\section{Simulating vortex ring collisions using the Biot-Savart approximation}
\subsection{Biot-Savart model and regularization}
\label{sec:BS}

In this section, we establish the conditions that lead to the emergence of the elliptical instability in vortex ring collisions, focusing on the range of Reynolds numbers in which this instability is observed. Our approach consists of modeling the rings using the Biot-Savart model~\cite{Moore:1972,Siggia:1985,PumirSig:1985} while accounting for the evolution of the core size, which is assumed to have a circular cross-section for all time. With the parameters given by this model, we determine the growth rate of the elliptical instability, as calculated in~\cite{LeDizes:2002}. 

Regularization of the Biot-Savart model is required due to the logarithmic divergence of the principal integral. Here, we use the regularization used in~\cite{Siggia:1985,PumirSig:1985}. 
Denoting $\sigma_i(t)$ as the core radius of each vortex filament, $i$, at time, $t$, we replace the Biot-Savart integral (up to a prefactor) by:

\begin{equation}
\frac{\partial \mathbf{r}_i(\theta,t)}{\partial t} = \sum_j \frac{\Gamma_j}{4 \pi}
\int_{{\rm filament}_j} d\theta' 
\frac{\frac{\partial \mathbf{r}_j}{\partial \theta'} \times (\mathbf{r}_i(\theta) - \mathbf{r}_j(\theta')) }{[ (\mathbf{r}_i(\theta) - \mathbf{r}_i (\theta') )^2 + \sigma_i(\theta)^2 + \sigma_j(\theta')^2 ]^{3/2} }.
\label{eq:BS_motion}
\end{equation}
In this configuration, we consider two filaments, with indices $i = 1 $ and $i = 2$. The centerline positions of the filaments with circulation, $\Gamma_i$, are given by $\mathbf{r}_i(\theta,t)$.
We start with an initially axisymmetric configuration, consisting of two counter-rotating vortex rings, perfectly aligned along the same central axis. We parametrize the centerlines of the two rings in the $(\hat{\mathbf{x}}, \hat{\mathbf{y}}, \hat{\mathbf{z}})$ directions by:

\begin{equation}
\mathbf{r}_1(\theta,t) = \begin{pmatrix} R(t) \cos(\theta) \\ R(t) \sin( \theta) \\-d(t)/2 \end{pmatrix}  ~~~ {\rm and } ~~~
\mathbf{r}_2(\theta,t) = \begin{pmatrix} R(t) \cos(\theta) \\ R(t) \sin( \theta) \\+d(t)/2 \end{pmatrix},
\label{eq:parametr}
\end{equation} 
where $R(t)$ is the vortex ring radius, and the azimuthal angle is given by $0 \le \theta \le 2 \pi$. The perpendicular distance between the two rings, $d(t)$, is taken to be positive, and the circulations are $\Gamma_1 = -\Gamma_2 = \Gamma > 0$. With these conventions, filament $1$, at $z = - d(t)/2$, 
moves upward, toward filament $2$ at $z = d(t)/2$.

In the following, as was the case in~\cite{Siggia:1985,PumirSig:1985}, we impose incompressibility, by enforcing that the total volume of the rings is conserved:
\begin{equation}
R(t) \sigma^2(t) = R_0\sigma_0^2.
\label{eq:sigma}
\end{equation}
The evolution equations reduce to two simple ordinary differential equations for $R(t)$ and $d(t)$, as explained in turn. With the parameterization proposed in Eq.~\eqref{eq:parametr}, an elementary calculation shows that the contribution of the filament $1$ to the velocity at the point $\mathbf{r}_1(\theta)$ reduces to a uniform velocity in the positive $z$-direction:

\begin{equation}
v_z^{1,s} = 
\frac{\Gamma}{4 \pi}
\int_{0}^{2\pi} d\theta' 
\frac{ R(t)^2 ( 1 - \cos( \theta - \theta') ) }
{[ 2 R(t)^2 (1 - \cos (\theta - \theta' ) )  + 2 \sigma^2 ]^{3/2} }.
\label{eq:vz_1s}
\end{equation}
The contribution of filament $2$ to the velocity of filament $1$ consists of a component in the radial direction:

\begin{equation}
v_r^{1,m} = 
\frac{\Gamma}{4 \pi}
\int_0^{2 \pi} d \theta' \frac{  d(t) R(t)  \cos( \theta - \theta' ) }
{[ 2 R(t)^2 (1 - \cos (\theta - \theta' ) ) + d(t)^2  + 2 \sigma^2 ]^{3/2} }
\label{eq:vr_1m}
\end{equation}
and a component in the vertical direction:

\begin{equation}
v_z^{1,m} = - \frac{\Gamma}{4 \pi}
\int_{0}^{2\pi} d\theta' 
\frac{ R(t)^2 ( 1 - \cos( \theta - \theta') ) }
{[ 2 R(t)^2 (1 - \cos (\theta - \theta' ) )  + d(t)^2 +  2 \sigma^2 ]^{3/2} }.
\label{eq:vz_1m}
\end{equation}
Hence, the evolution equation of the vortex ring radius, $R(t)$ reduces to:

\begin{equation}
\dot{R}(t)  = v_r^{1,m}  = 
\frac{\Gamma d(t) }{2 \pi R(t)^2} \Psi[ (d(t)^2+ 2\sigma^2)/(2 R(t)^2) ],
\label{eq:R_dt}
\end{equation}
where $\Psi$ is defined as:
\begin{equation}
\Psi(X^2) = \int_0^{2 \pi} d \theta' \frac{  \cos( \theta - \theta' ) }
{[ 2 (1 - \cos (\theta - \theta' ) ) + 2 X^2 ]^{3/2} }.
\label{eq:Psi}
\end{equation}
Similarly, the evolution equation of the spacing between the rings, $d(t)$, is given by:

\begin{equation}
\dot{d}(t) = v_z^{1,s} + v_z^{1,m}
= \frac{\Gamma }{ 2 \pi R(t)} \bigg( 
\Phi[ ( d(t)^2 + \sigma^2 )/(2 R(t)^2) ] 
- 
\Phi[\sigma^2/R(t)^2 ] 
\bigg),
\label{eq:d_dt}
\end{equation}
where $\Phi$ is defined as:
\begin{equation}
\Phi(X^2) = \int_0^{2 \pi} d \theta' 
\frac{ ( 1 -  \cos( \theta - \theta' ) ) }
{[ 2 (1 - \cos (\theta - \theta' ) ) + 2 X^2 ]^{3/2} }.
\label{eq:Phi}
\end{equation}
It is a simple matter to compute asymptotic expressions for the functions $\Phi$ and $\Psi$ when $X^2 \rightarrow 0$ or $X^2 \rightarrow \infty$. 
To determine the evolution of $R(t) $ and $d(t)$, we numerically integrate Eq.~\eqref{eq:sigma}, Eq.~\eqref{eq:R_dt} and Eq.~\eqref{eq:d_dt}.

\subsection{Dynamic evolution of the vortex rings}
\label{sec:num_results}

Eq.~\eqref{eq:sigma}, Eq.~\eqref{eq:R_dt}, and Eq.~\eqref{eq:d_dt} are used to compare the radial growth of the colliding vortex rings from the Biot-Savart model with the experimental data in the inset of Fig. 1(B) in the main text. This model agrees well with the mean radial growth of the rings prior to breaking down, at which point the assumptions of Biot-Savart are clearly violated.
Here, we further compare the model against direct numerical simulations which describe the head-on collision of two vortex rings with the following initial conditions: $R(0) = R_0 $, $d(0) = 2.5R_0$, and $\sigma(0) = 0.1R_0$. This choice of parameters allows for the direct comparison of the model with the DNS, starting from the same initial configuration. The DNS collision, where Re$_{\Gamma} = 4500$, was already presented in Fig. 2(B,D) of the main text and in Movie~\ref{mov:sim_rings_dye}.

\begin{figure}[H]
\centering
\includegraphics[width=\textwidth]{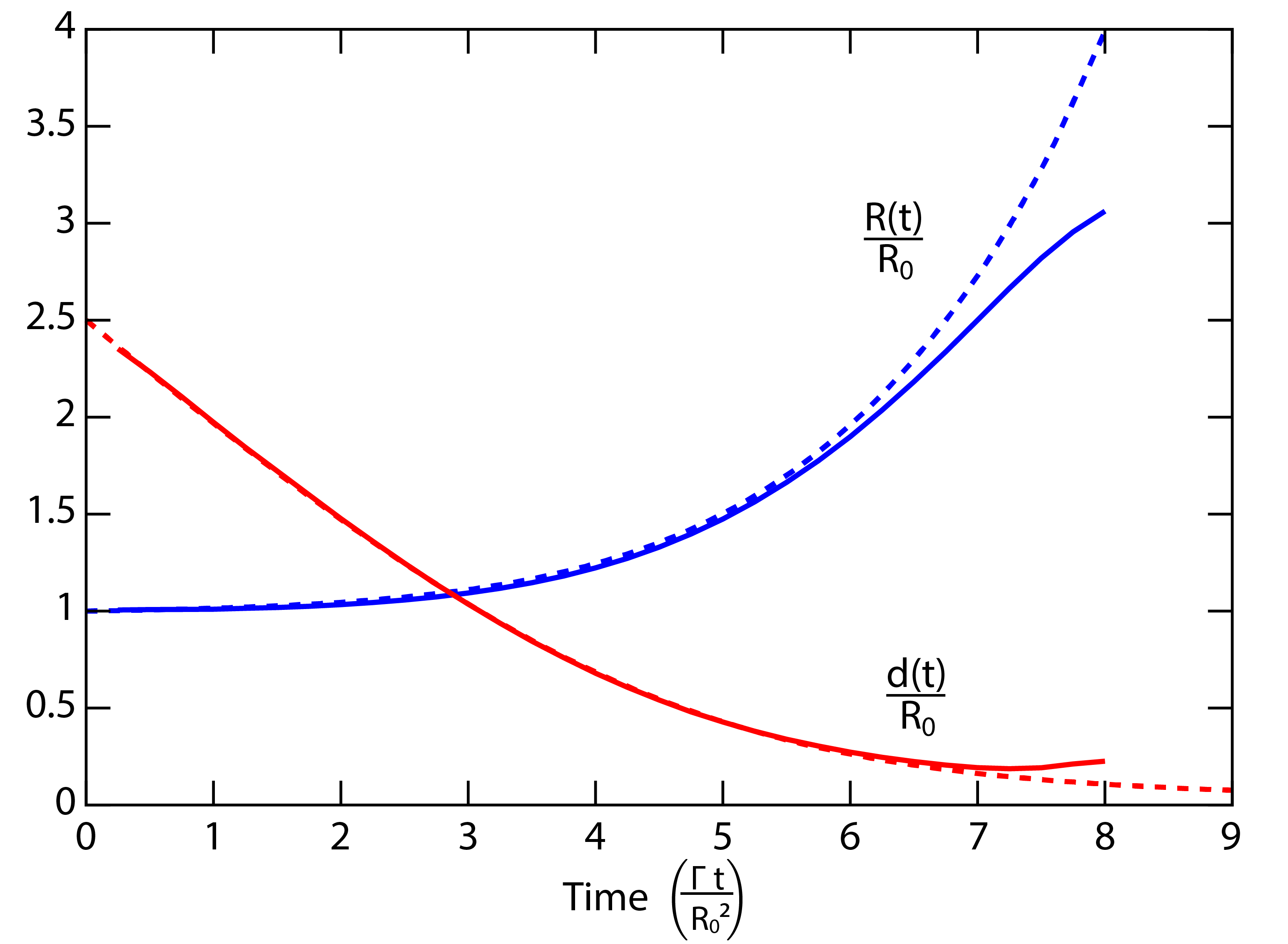}
\caption{
Comparison between the evolution of the perpendicular spacing between the cores, $d(t)$, and the vortex ring radius, $R(t)$, obtained from DNS of the Navier-Stokes equations where Re$_\Gamma = 4500$ (full lines) and from the Biot-Savart model (dashed lines). For both configurations, $d(0) =  2.5R_0$ and $\sigma(0) = 0.1R_0$.
}
\label{fig:compare_bs_ns}
\end{figure}

Fig.~\ref{fig:compare_bs_ns} shows the evolution of the vortex ring radius, $R(t)$, (blue curves) and the perpendicular distance between the filaments, $d(t)$, (red curves). The solution of the Biot-Savart model is shown with dashed lines, and the full lines correspond to the same configuration evaluated numerically by solving the Navier-Stokes equations.
As the two rings initially approach one another, the distance between them decreases linearly. However, in both the Biot-Savart model and the DNS, the rings reach a minimum distance on the order of the core size before breaking down. 
During the late stage of the collision, the rings grow radially outwards; however, the growth of the ring radius predicted by the Biot-Savart model significantly exceeds that of the collision obtained from the DNS because the core dynamics--and hence the breakdown--is not captured by Biot-Savart model.

\subsection{ Elliptical instability }
\label{sec:instab}

Here we examine the the onset of the elliptical instability when the vortex rings collide head-on. We are primarily interested in the growth of short-wavelength modes, characteristic of the elliptical instability~\cite{Leweke:1998,Leweke:2016}. 
For this reason, we neglect the curvature of the rings and approximate the filaments with a pair of straight antiparallel vortices, located at $z = \pm d(t)/2$, with circulations $\pm \Gamma$, and each with a core radius $\sigma(t)$. The values of $d(t)$ and $\sigma(t)$ are determined from the solutions of the Biot-Savart model.

The onset of the elliptical instability results in the development of perturbations with a wavelength on the order of the core size~\cite{Leweke:2016}.
Our analysis is based on the work of LeDizes, who examined the elliptical instability in the same flow configuration ~\cite{LeDizes:2002,LeDizes:2002_corr}. LeDizes derived the following equation for the growth rate, $\gamma$, of an infinitesimal perturbation of the vortex cores with a longitudinal wavenumber $k_z$:

\begin{equation}
\gamma = \frac{\Gamma}{8 \pi d^2 } 
\sqrt{ \left(\frac{3}{4}\right)^4 K_{NL}(0)^2 - \frac{64 d^4}{\sigma^4} 
\left(\frac{1}{2} - \cos(\zeta^{(m)}) \right)^2 } - \frac{8 \pi k_z^2 d^2 }{\text{Re}_{\Gamma} \cos(\zeta^{(m)})   },
\label{eq:instab_ell}
\end{equation}

where $K_{NL}(0) = 1.5 + 0.038 \times 0.16^{-9/5} \approx 2.52$, and
$\cos (\zeta^{(m)}) = \big( \frac{1}{2} - \frac{(2.26 + 1.69 m) - k_z \sigma}{14.8 + 9m } \big)$.

In the expression for $\cos (\zeta^{(m)})$, $m$ is an integer that distinguishes between several branches of solutions. We only examine the most unstable branch, corresponding to $m = 0$. To estimate the growth rate of the elliptical instability, we consider an azimuthal perturbation mode, $n$, along the rings. This corresponds to a wavenumber $q_n(t) = n/R(t)$. The growth of the radius, $R(t)$, implies that the wavenumber $q_n(t)$ decreases as time progresses for a fixed value of $n$. Note that when neglecting the curvature of the colliding rings, $k_z = q_n(t)$.

\begin{figure}[hb!]
\centering
\includegraphics[width=\textwidth]{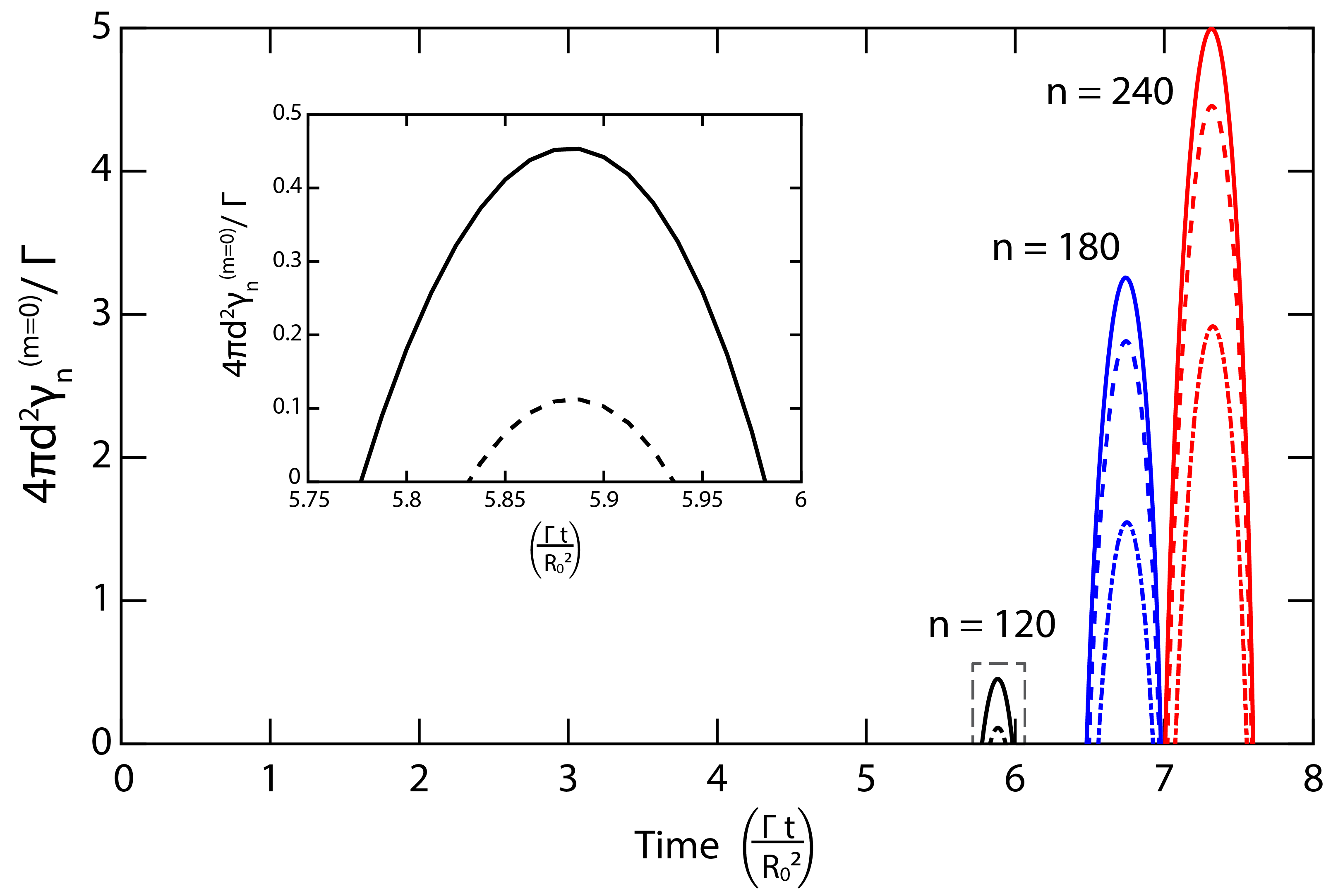}
\caption{
Onset of the elliptical instability for colliding vortex rings. The normalized growth rate, $\gamma_{n}$, of three azimuthal perturbation modes, $n$, is evaluated for the head-on collision of two vortex rings using Eq.~\eqref{eq:instab_ell}. (inset) zoomed-in view of the plot of the $n=120$ mode indicated by the dashed gray box.
The full lines, dashed lines, and dot-dashed lines correspond to Re$_\Gamma = 5000, 4500, \text{and} ~3500$, respectively. The values of $R(t)$ and $d(t)$ are calculated from the Biot-Savart model, shown by the dashed lines in Fig.~\ref{fig:compare_bs_ns} above. 
}
\label{fig:growth_rate}
\end{figure}

We evaluate the growth rate of the elliptical instability, $\gamma_n^{(m=0)}$, as a function of time for $n = 120, 180,$ and $240$, focusing on only the times when the growth rate is positive, as shown in Fig.~\ref{fig:growth_rate}.
These values of $n$ were selected because they correspond to wavelengths that are on the order of the core radius, $\sigma(t)$, when the instability begins.
We examine how the growth rates of these modes evolve when Re$_\Gamma = 3500$ (dotted-dashed lines), Re$_\Gamma = 4500$ (dashed lines) and Re$_\Gamma = 5000$ (full lines). 
The elliptical instability develops at each value of $n$  only over a short period of time. The magnitude of the growth rate $\gamma_n^{(m=0)}$ increases as the Reynolds number increases.

The Biot-Savart model--Eq.~\eqref{eq:sigma}, Eq.~\eqref{eq:R_dt}, and Eq.~\eqref{eq:d_dt}--provides a semi-quantitative description of the solutions of the Navier-Stokes equations when the $n=120$ mode becomes unstable, as shown in the inset of Fig.~\ref{fig:growth_rate}. At this time, $\Gamma t/ R_0^2=5.9$, the unstable wavelength is $\lambda \approx 2 \pi R(t)/120 \approx 2 \sigma(t)$. The model therefore predicts that perturbations with a wavelength on the order of the core size are unstable due to the elliptical instability mechanism when Re$_\Gamma = 4500$, but are stable for Re$_\Gamma = 3500$, in qualitative agreement with our findings. The elliptical instability is still triggered for the Re$_\Gamma = 3500$ configuration, albeit at later times.
The model also demonstrates that the onset of the elliptical instability for Re$_\Gamma = 4500$ occurs when the spacing between the rings, $d(t)$, reaches the minimum threshold on the order of the initial core thickness, $2\sigma_0 = 0.2R_0$, as shown in Fig.~\ref{fig:compare_bs_ns}. This is consistent with the experimental and DNS results shown in Fig. 2(C-D) in the main text.

\section{Nonlinear development of the elliptical instability }
\label{sec:transverse_filaments}

\subsection{ Formation of secondary vortex filaments }
Following the development of antisymmetric perturbations that result from the elliptical instability, an array of secondary vortex filaments spontaneously forms perpendicular to the original vortex cores. This has been directly observed experimentally and numerically for vortex ring collisions, as detailed in the main text. The same flow structures emerge via the elliptical instability during the interaction between two antiparallel vortex tubes~\cite{Leweke:1998,Laporte:2000}. 
This observation seems at first surprising, as it indicates the emergence of a component of circulation in the plane separating the two vortices, where the vorticity is initially zero. If the midplane separating the two vortices (i.e. the $z=0$ plane) were a plane of symmetry--as is the case in many studies examining the Crow instability~\cite{Crow:1970} of interacting vortex tubes~\cite{Pumir:1990,Kerr:1993}--the vorticity along this plane would remain zero at all times. 
The antisymmetric nature of the perturbed vortex cores, shown in Fig. 2(A-B) of the main text, however, allows for the development of a non-zero circulation in the collision plane. Here, we examine how a significant component of the circulation can accumulate on the plane $z = 0$, referred to here as the plane of reflection, as shown in Fig. 4(C-D) in the main text.

\begin{figure}[b!]
\centering
\includegraphics[width=0.92\textwidth]{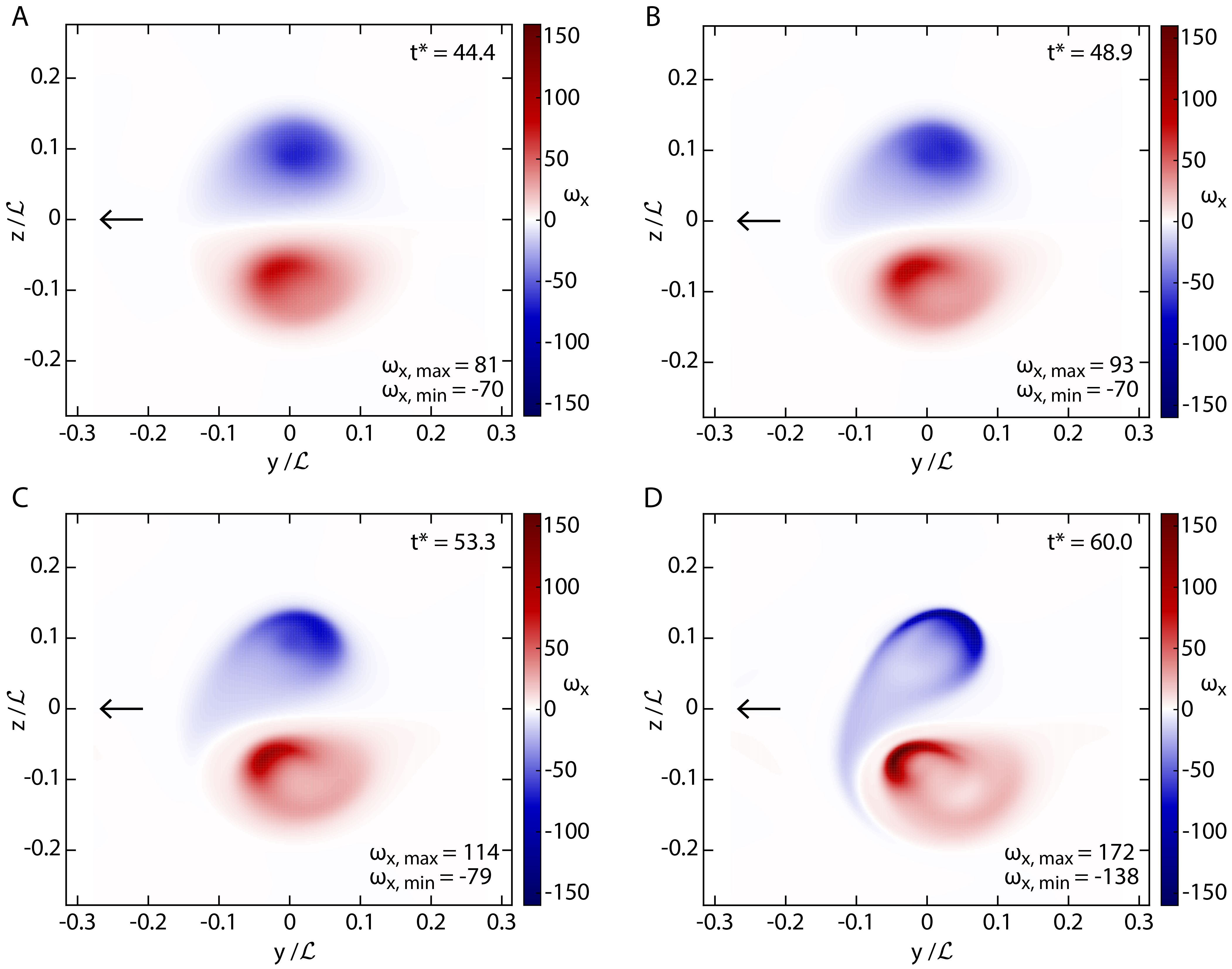}
\caption{
Formation of a secondary vortex filament. Temporal evolution of the axial vorticity distribution, $\omega_x(y,z)$, along a fixed cross section at $x/\mathcal{L} = 0.48$ for DNS of counter-rotating vortex tubes where Re$_\Gamma = 4500$. The black arrows indicate the propagation direction, though a horizontal offset is applied at each time to keep the vortices in the center of the domain. (A) The vorticity from the two vortex tubes is initially separated along the ($z=0$) reflection plane. (B) As the perturbations develop, the lower core migrates to the leading direction and the upper core migrates to the trailing end. (C) The cores flatten into sheet-like structures, and vorticity from the upper vortex is advected down to the lower vortex. (D) The cores contract into highly curved, sheetlike structures where the vorticity is concentrated. The advected vorticity forms a secondary vortex filament across the ($z=0$) reflection plane. Note: $\sigma = 0.06 \mathcal{L}$, $b = 2.5 \sigma$, and $t^* = \Gamma t / b^2$.  
}
\label{fig:transv_tubes}
\end{figure}

We examine the same direct numerical simulation presented in Fig. 4 in the main text and Movie~\ref{mov:tubes_Re_4500}, consisting of two, initially parallel, counter-rotating vortex tubes with Re$_\Gamma = 4500$. We examine the evolution of the normal vorticity component,  $\omega_x$, along a fixed axial cross-section of the tubes, as shown in Fig.~\ref{fig:transv_tubes}. 
The upper vortex rotates in the clockwise direction, and the lower vortex rotates in the opposite direction, resulting in their propagation in the negative $y-$direction, indicated by the black arrows.
Initially, the vorticity of the tubes largely remains on the respective sides of the plane of reflection, as shown in Fig.~\ref{fig:transv_tubes}(A). However, once the tubes become perturbed, the centroids of the vortex cores deform such that the lower core moves forward in the propagation direction and the upper core moves backward, as shown in Fig.~\ref{fig:transv_tubes}(B-D). This contrary motion of the cores illustrates the antisymmetric structure of the perturbations, as this particular cross section is located along an anti-node of the pair. While the amplitudes of the perturbations grow, the vorticity distributions of the cores contract and amplify into flattened, sheet-like structures, as shown in Fig.~\ref{fig:transv_tubes}(D) and Fig. 4(A) in the main text. The curvatures of the deformed cores are of opposite sign; the leading vortex core is curved toward the propagation direction and vice-versa for the trailing core. 
Additionally, the lower core, which is deflected toward the leading edge of the vortex pair, has a higher curvature than that of the upper core.
The tendency of the kinked, perturbed cores to locally flatten into vortex sheets results from the stretching field generated by each filament, as characterized by the Biot-Savart model. An elementary calculation predicts that, on the outer side of a kinked filament, the vorticity is stretched and grows, while on the inner side, the vorticity decreases (see Fig. 4 of~\cite{Pumir:1990}).
The tendency of the perturbed vortex cores to flatten into sheets is therefore the result of the dynamics of kinked vortex tubes.

\begin{figure}[htb]
\centering
\includegraphics[width=\textwidth]{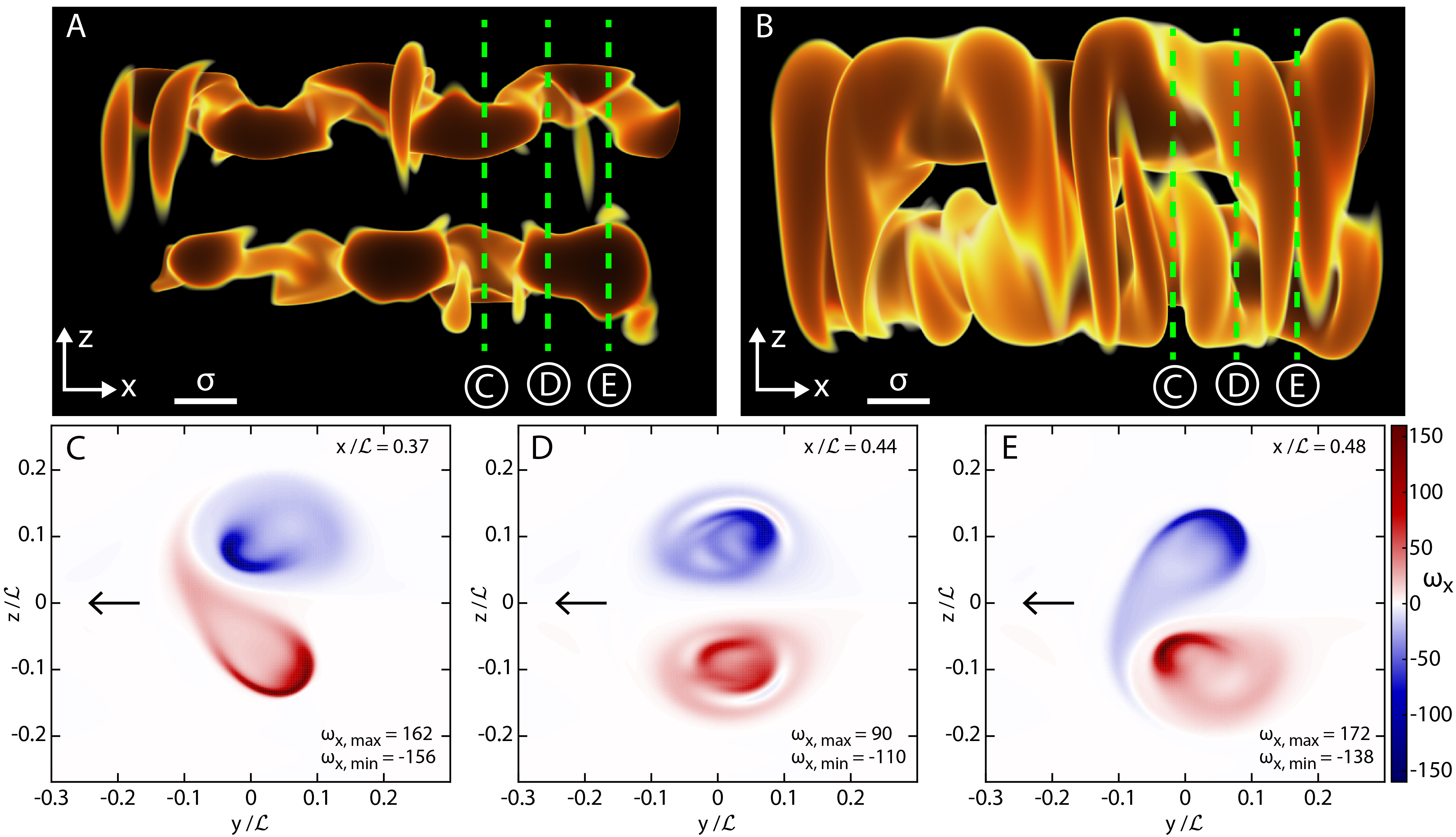}
\caption{
Alternating structure of secondary filaments. DNS of counter-rotating vortex tubes where Re$_{\Gamma}$=4500 at a fixed time, $t^* = \Gamma t / b^2 = 60.0$. (A-B) 3D vorticity modulus of the simulated flow and (C-E) distribution of axial vorticity, $\omega_x(y,z)$, along cross sections of the tubes indicated by the dashed green lines. The black arrows indicate the propagation direction of the vortex pair. (A) The cores flatten into vortex sheets at the tips of each perturbation ($0.092 \leq |\omega|/|\omega|_{\text{max}} \leq 0.122$). (B) The edges of the each flattened perturbation roll up into pairs of secondary filaments ($0.031 \leq |\omega|/|\omega|_{\text{max}} \leq 0.053$). Due to the antisymmetric structure of the perturbations, the orientation of the secondary filament pairs alternates periodically. (C) At anti-nodes where the top vortex core leads, the secondary filament is stretched up from the lower vortex tube. (D) At nodes, neither vortex core leads and no secondary filaments form. (E) At anti-nodes where the bottom vortex core leads, the secondary filament is stretched down from the upper vortex tube. Note: $\sigma = 0.06 \mathcal{L}$ and and $b = 2.5\sigma$.
}
\label{fig:contours_vrt_b}
\end{figure}

Because the perturbations of the vortex cores are periodic, the relative positions of the cores revert every half-wavelength along the axial direction of the tubes, as shown in Fig.~\ref{fig:contours_vrt_b}. The 3D visualization of the vorticity modulus, $|\omega|$, shows how the peaks of the perturbed cores flatten into mushroom-cap structures, as shown in Fig.~\ref{fig:contours_vrt_b}(A). By lowering the threshold of the vorticity modulus, as shown in Fig.~\ref{fig:contours_vrt_b}(B), one can visualize how the edges of the flattened cores roll up into two secondary vortex filaments that are pulled toward the opposite vortex tube.     
Moving along the axial direction, the leading vortex switches from the top core (Fig.~\ref{fig:contours_vrt_b}(C)) to the bottom core (Fig.~\ref{fig:contours_vrt_b}(E)). At the nodes of the perturbations, the cores are aligned with each other, as shown in Fig.~\ref{fig:contours_vrt_b}(D).  
Notably, the inherent asymmetry of the offset vortices causes the highly curved, leading core to locally advect the low-vorticity region of the trailing core around itself, as previously shown in Fig.~\ref{fig:transv_tubes}. 
This shedding of vorticity repeats along each anti-note peak of the perturbed cores, leading to an alternating array of perpendicular secondary filaments that traverse the plane of reflection, as shown in Fig.~\ref{fig:contours_vrt_b}(B).
The alternation of pairs of secondary filaments accounts for the interdigitation of the colliding vortices visualized with dye both experimentally and numerically in the main text.
The counter-rotating structure of adjacent secondary filaments, as shown in Fig. 4(D) in the main text, results from two sources.
First, the edges of each flattened perturbation roll up into a pair of secondary filaments that counter-rotate relative to one another. 
Additionally, the alternating orientation of the secondary vortex pairs cause filaments formed from adjacent perturbations to counter-rotate with each other.
\subsection{ Transfer of circulation }

\begin{figure}[H]
\centering
\includegraphics[width=\textwidth]{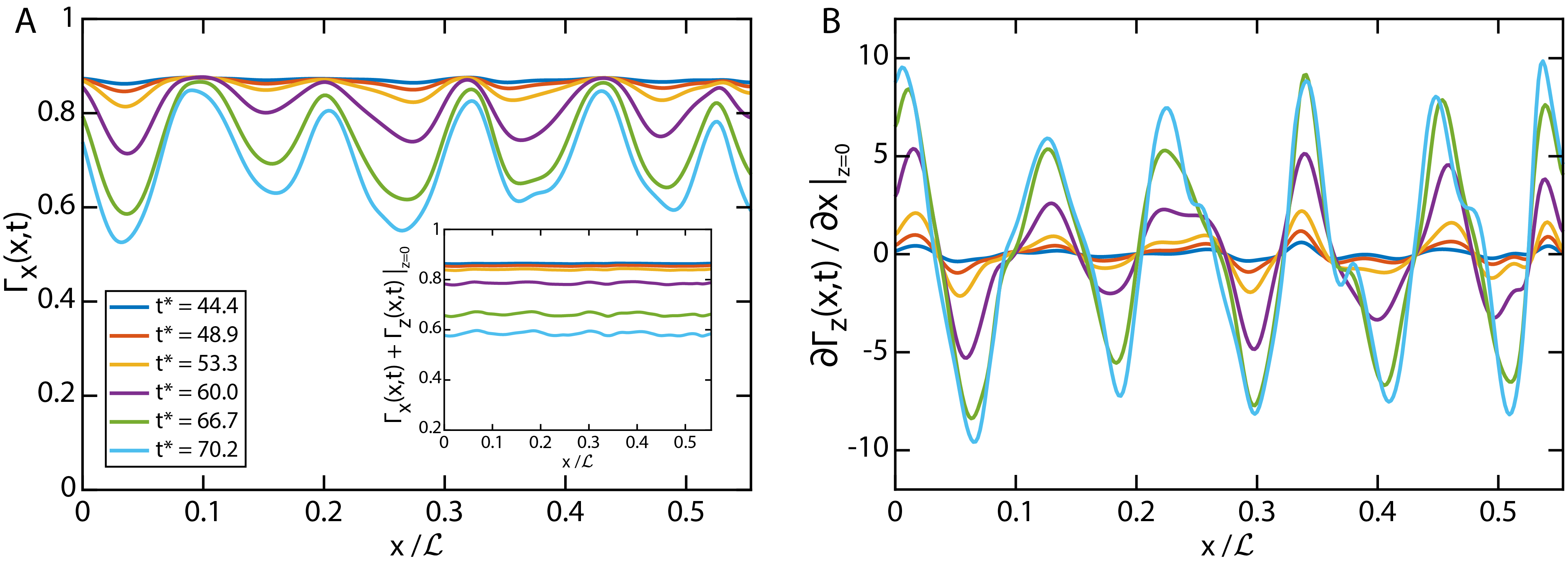}
\caption{Evolution of circulation. DNS of counter-rotating vortex tubes where Re$_{\Gamma} = 4500$. (A) Evolution of the axial component of circulation, $\Gamma_x(x,t)$, in the lower half plane $(y,z)$ for $z < 0$. (inset) Evolution of the axial and transverse components of circulation, $\Gamma_x(x,t) + \Gamma_z(x,t)$. (B) Evolution of the derivative of the transverse component of circulation, $\partial \Gamma_z/\partial x$ along the reflection plane $(z=0)$.
}
\label{fig:circ_x}
\end{figure}

The dynamics of the thin secondary filaments, transported across the reflection plane ($z=0$), vary along the axial direction. This implies that the axial component of the circulation in the half-plane,

\begin{equation}
\Gamma_x(x,t) \Big|_{z \to -\infty}^{z = 0} \equiv \int_{-\infty}^\infty dy \int_{-\infty}^0 dz \, \omega_x(x,y,z,t) 
\label{eq:circ_x}
\end{equation} 
varies, both as a function of time and of $x$. By integrating the vorticity distribution on the lower half plane at several times, we find that the axial component of circulation varies periodically with $x$, as shown in Fig.~\ref{fig:circ_x}(A). This demonstrates how the formation and stretching of the secondary filaments develop a periodic transfer of vorticity from the original tubes through the reflection plane. Once the secondary filaments are fully formed after $t^* \gtrsim 65$, however, the variations in the axial component of circulation saturate.

Concomitantly, as the axial component of circulation is drained from the lower vortex tube, the secondary filaments traverse through the reflection plane, $z=0$. 
The corresponding flux of vorticity in the transverse direction is quantified by computing 
the transverse component of circulation $\Gamma_z(x,t)$ along the reflection plane on $ [0, x]$: 
\begin{equation}
\Gamma_z (x,t) \equiv \int_0^x dx' \int_{-\infty}^{+\infty} dy' \, 
\omega_z(x',y',z = 0) 
\label{eq:Gam_z}
\end{equation}

The derivative of $\Gamma_z(x,t) $ with respect to $x$ reduces Eq.~\eqref{eq:Gam_z} into the normalized sum of the transverse vorticity, $\omega_z$, along a line located at the axial position, $x$, on the reflection plane, as shown in Fig.~\ref{fig:circ_x}(B) at various times. 
The sinusoidal nature of $\partial\Gamma_z(x,t)/\partial x |_{z=0}$--about a mean value of zero--along the axial direction showcases the counter-rotating structure of adjacent secondary filaments, as shown in Fig. 4(D) in the main text.

An important relation between $\Gamma_z(x,t)$ and $\Gamma_x(x,t)$ results from the observation that $\nabla \cdot \mathbf{\omega} = 0$; that is, the flux of vorticity a through a closed surface limiting a finite volume of fluid is conserved. 
We define the half-plane ${\cal P}_\alpha$ by the conditions $x = \alpha$ and $z \le \alpha$, and the band ${\cal Q}_{\alpha \beta}$ for $\alpha \le \beta$ by $z = 0$ and $\alpha \le x \le \beta$.
Additionally, we use the property that $\nabla \cdot \mathbf{\omega} = 0$ to the domain limited by ${\cal P}_\alpha$, ${\cal Q}_{\alpha \beta}$, and ${\cal P}_\beta$, with $\alpha \le \beta$. 
An elementary calculation shows that the flux of $\mathbf{\omega}$ on this domain reduces to $ \big(\Gamma_x(\beta,t) + \Gamma_z(\beta,t) \big) - \big(\Gamma_x(\alpha,t) + \Gamma_z(\alpha,t) \big) $. 
Thus, the condition that the flux of vorticity is zero imposes that $\Gamma_x(x,t) + \Gamma_z(x,t) $ does not dependent on $x$. 
We find that our numerical results satisfy this conservation relation, as shown in the inset of  Fig.~\ref{fig:circ_x}.

Physically, the relation $\partial_x \big( \Gamma_x(x,t) + \Gamma_z(x,t) \big)=0$ imposes that the axial variations of $\Gamma_x(x,t)$, clearly visible in Fig.~\ref{fig:circ_x}(A), necessitate the variations of the flux $\Gamma_z(x,t)$, which are proportional to the transverse circulation in the plane $z = 0$. 
That is, because the circulation of the system is conserved--aside from viscous dissipation--any axial  circulation lost from the vortex tubes is redirected to the secondary filaments through the transfer of transverse circulation. 
The fluctuations of $\Gamma_x(x,t)$ correspond to $\approx 0.25 \Gamma_0$, implying that in the configuration studied here, the circulation of each transverse secondary filament is approximately $1/4$ of the axial component of circulation in the original tubes.

\section{Interactions of secondary vortex filaments}

\begin{figure}[H]
\centering
\includegraphics[width=0.9\textwidth]{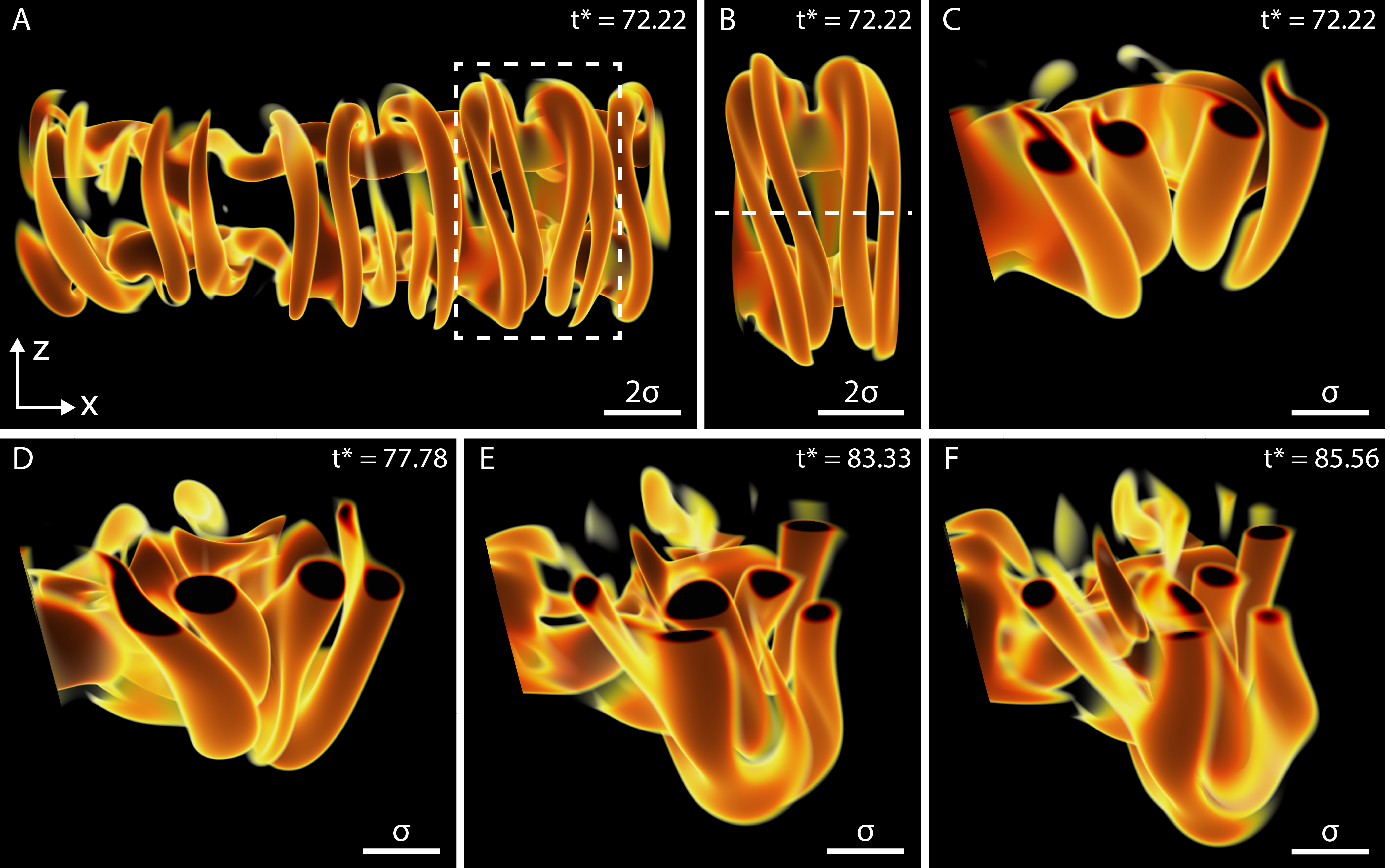}
\caption{Interactions of secondary vortex filaments. Vorticity modulus of a simulated vortex tube interaction where Re$_{\Gamma} = 3500$. (A) Array of secondary filaments formed during the late-stage evolution of the elliptical instability. (B) Zoom-in view of two pairs of secondary filaments indicated by the dashed box. (C) Cross-sectional view of the secondary filaments through the plane of reflection $(z=0)$, indicated by the dashed line in (B). Neighboring filaments counter-rotate and interact with each other. (D) Interacting secondary filaments deform from the mutual strain and one of the filaments locally flattens into a vortex sheet. (E) The flattened vortex sheet splits into two smaller tertiary vortex filaments. (F) The newly-formed tertiary filaments unravel the secondary filament. For (A-F), the vorticty threshold is $0.122 \leq |\omega|/|\omega|_{\text{max}} \leq 0.206$, where $|\omega|_{\text{max}}$ is the maximum vorticity modulus for the entire simulation. Note: $\sigma = 0.06 \mathcal{L}$, $b = 2.5 \sigma$, and $t^* = \Gamma t / b^2$. }
\label{fig:vortex_sheets}
\end{figure}

Through the creation of secondary filaments, the elliptical instability provides a mechanism by which smaller generations of counter-rotating vortex filaments form and interact to generate small-scale flow structures. 
During the evolution of the elliptical instability, the development of antisymmetric perturbations in the cores leads to the formation of an array of counter-rotating secondary filaments, as shown numerically for a typical example in Fig.~\ref{fig:vortex_sheets}(A), Movie~\ref{mov:tubes_Re_3500}, and Movie~\ref{mov:sheets}, where Re$_\Gamma = 3500$.    
Neighboring secondary filaments interact with one another in the same manner as the initial vortex tubes. Because the secondary filaments are smaller and have a lower circulation than the original vortex tubes, they can be viewed as having a lower effective Reynolds number. 
These adjacent secondary filaments align with one another in pairs, as shown in Fig.~\ref{fig:vortex_sheets}(B-C). 
Due to their counter-rotation, the filaments exert large strains on each other, which causes one of the secondary filaments to flatten into a vortex sheet, as shown in Fig.~\ref{fig:vortex_sheets}(D). Upon further stretching, this vortex sheet splits into two smaller tertiary vortex filaments, as shown in Fig.~\ref{fig:vortex_sheets}(E-F).
This breakdown mechanism in which counter-rotating vortex filaments interact, flatten into vortex sheets, and split into even smaller generations of vortices has been previously observed in vortex ring collisions~\cite{McKeown:2018} and is attributed to the late-stage development of the Crow instability~\cite{Crow:1970}. 
This instability dominates during the interaction of vortices at low-Reynolds numbers, like that of the secondary vortex filament pair. 

As shown previously, only a fraction of the initial circulation is transferred to the secondary filaments. Even though the elliptical instability fully develops during the Re$_\Gamma = 3500$ configuration, the effective Reynolds number of the interacting secondary filaments is not sufficient enough for the elliptical instability to develop again. Instead, the Crow instability dominates during this second iteration, and the secondary vortices flatten into vortex sheets and split into smaller tertiary filaments. 
In order for the secondary filaments to, themselves, interact through the elliptical instability and form a tertiary generation of perpendicular filaments--as shown in Fig. 5 in the main text and in Movie~\ref{mov:cascade}--the initial counter-rotating vortex tubes must have a higher Reynolds number.

\section{Analysis of the transfer of energy in a turbulent flow }
\label{sec:analysis_transfer}

This section examines the derivation and meaning of the shell-to-shell energy transfer spectrum, $T(k,t)$, introduced in the main text and plotted in Fig. 5(G).
A typical method for characterizing a turbulent flow, which encompasses of a wide range of excited scales of motion, is to examine evolution of the the energy spectra in Fourier space. 
This energy spectrum is designated by the term $E(k,t)$, such that $E(k,t)~dk$ is the amount of kinetic energy at time $t$, in a shell in wavenumber space between $k$ and $k + dk$. 
In the absence of forcing, and in the simplified case of a homogeneous isotropic flow, one can derive from the Navier-Stokes equations the following energy balance~\cite{Pope:2000,Lin:1947}:

\begin{equation}
\frac{\partial E(k,t)}{\partial t} + T(k,t) = - 2 \nu k^2 E(k,t).
\label{eq:Lin}
\end{equation}

The terms in this equation state that for a given wavenumber, $k$, and at any time, t, the rate of change of the energy of that mode plus the rate of energy transferred to or from that wavenumber via other modes is balanced by the viscous dissipation of that mode.
In this equation, $T(k,t)$ originates from the nonlinear, advective term in the Navier-Stokes equations:

\begin{equation}
T(k,t) \, dk = \sum_{k \le | \mathbf{k} | \le k + dk } 
\Re{ \big( \overline{ (\mathbf{u} \cdot \nabla ) \mathbf{u} } (\mathbf{k} )
\cdot \overline{ \mathbf{u}} (- \mathbf{k} ) \bigr) }. 
\label{eq:Pi}
\end{equation}
While Eq.~\eqref{eq:Lin} is often used in a context of fluid turbulence, it can be applied to any solution of the Navier-Stokes equations. 
In our direct numerical simulations, $T(k,t)$ is calculated by applying a discrete Fourier transform to both our solved flow field, $\mathbf{u}(\mathbf{x},t)$ and the nonlinear term,
$( \mathbf{u} \cdot \nabla ) \mathbf{u} (\mathbf{x}, t)$, which are then applied to Eq.~\eqref{eq:Pi}.

\section{Emergence of turbulence from the elliptical instability with increasing Reynolds number \label{sec:Tubes_Across_Re}}

\subsection{Dissipation rate evolution and energy spectra \label{subsec:Spectra_Across_Re}}

\begin{figure}
\centering
\includegraphics[width=0.9\textwidth]{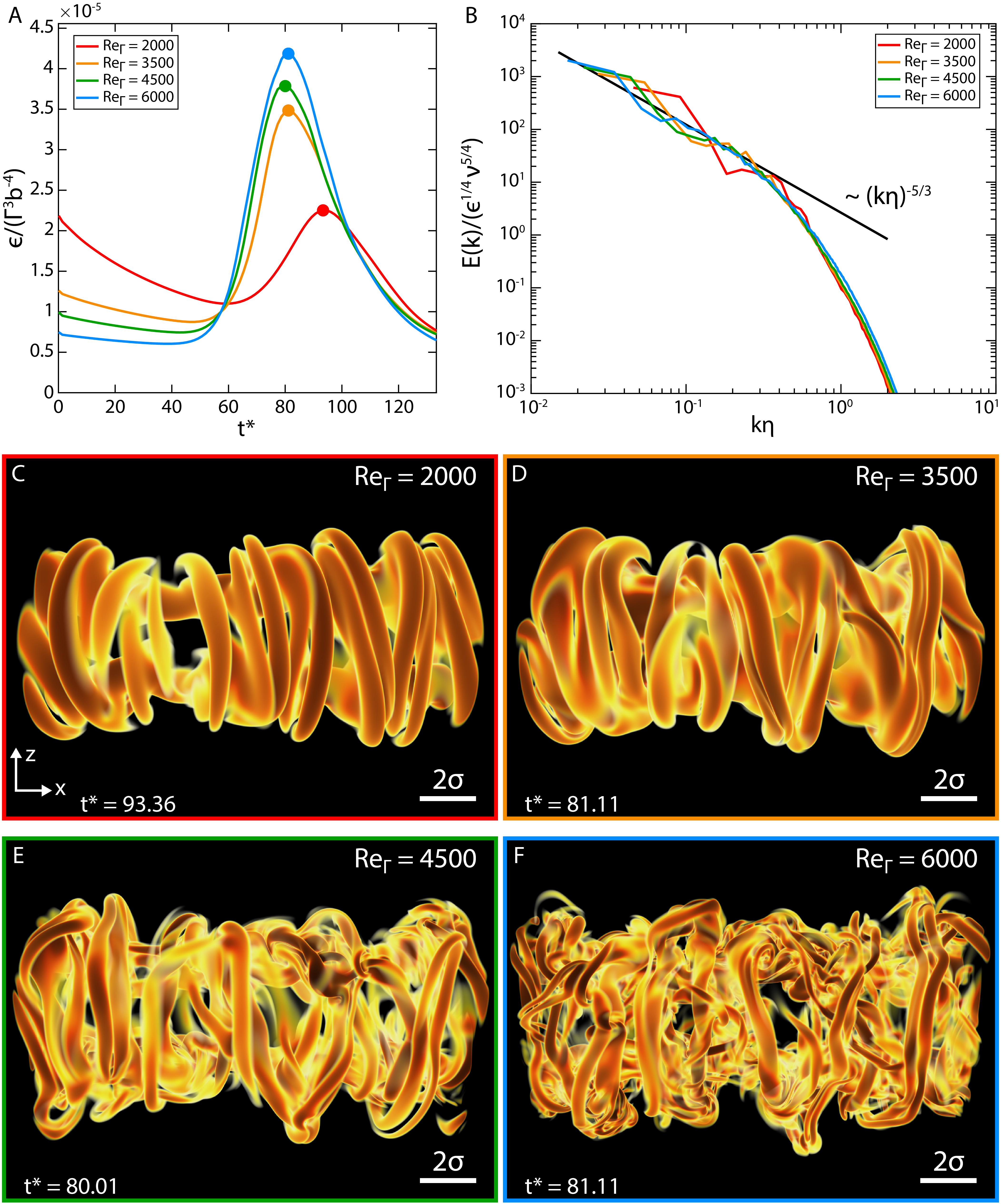}
\caption{Dissipation rate evolution and energy spectra for simulated vortex tube interactions at various Reynolds numbers. (A) Evolution of normalized kinetic energy dissipation rate. The markers indicate the maximum energy dissipation rate. (B) Normalized kinetic energy spectra at the peak dissipation rate, where the black line corresponds to Kolmogorov scaling. (C-F) Snapshots of the 3D vorticity modulus at the times corresponding to the maximum dissipation rate for each Reynolds number. The vorticity thresholds are $0.153 \leq |\omega|/|\omega|_{\text{max}} \leq 0.305$ for (C) and $0.061 \leq |\omega|/|\omega|_{\text{max}} \leq 0.183$ for (D-F).
For each simulation, $\sigma = 0.06\mathcal{L}$, $b = 2.5\sigma$, Re$_{\Gamma} = \Gamma/\nu$, and $t^* = \Gamma t / b^2$. 
}
\label{fig:epsilon_energy_spectra}
\end{figure}

Direct numerical simulations of the interacting, counter-rotating vortex tubes are performed at a range of Reynolds numbers to examine by what mechanism the onset of the elliptical instability leads to the development of turbulence. At each Reynolds number, the energy dissipation rate, $\epsilon$, qualitatively follows the same temporal evolution, as shown in Fig.~\ref{fig:epsilon_energy_spectra}(A).
The coherent vortex tubes initially interact, and the rapid increase in $\epsilon$ is initiated by the onset of the elliptical instability at each Reynolds number, as shown in Fig. 5 in the main text. 
The dissipation rate is maximized during the late-stage of the elliptical instability, in which the secondary filaments interact with each other and the remnants of the original primary vortex cores. As the Reynolds number is increased, the maximum dissipation rate increases. Because the viscous dissipation of energy in the flow primarily occurs on the smallest scales of the system, this behavior indicates that the high-Reynolds number configurations more effectively convey energy into small-scale flow structures.

Additionally, we examine the normalized kinetic energy spectra, $E(k)/(\eta^{\frac{1}{4}} \nu^{\frac{5}{4}})$, when the dissipation rate is maximized for each Reynolds number, as shown in Fig.~\ref{fig:epsilon_energy_spectra}(B). Each of the energy spectra generally follow the $\sim(k \eta)^{-5/3}$ Kolmogorov scaling of turbulence, where $\eta$ is the dissipative length scale~\cite{Kolmogorov:1941}. 
Notably, the agreement of the simulated energy spectra with this turbulent scaling improves for simulations that are carried out at higher Reynolds numbers. This is because the inertial range of the breakdown is more developed at higher Reynolds numbers--i.e. there is a larger range of scales over which Kolmogorov's axioms for turbulence are valid~\cite{Kolmogorov:1941}. 
The emergence of this multi-scale turbulent behavior is encapsulated by the snapshots in Fig.~\ref{fig:epsilon_energy_spectra}(C-F) which show the vorticity modulus of the interacting tubes at each Reynolds number when $\epsilon$ is maximized. 

In each configuration, the elliptical instability is fully developed at the peak dissipation rate, as an array of perpendicular secondary filaments bridges the gap between the original vortex tubes. 
These stretched, counter-rotating secondary filaments interact with each other and the remnants of the original tubes through different means at each Reynolds number. For the Re$_{\Gamma} = 2000$ configuration, the secondary filaments do not have sufficient circulation to interact with one another and break down further, as viscous dissipation sets in (see Movie~\ref{mov:tubes_Re_2000}). When the Reynolds number is raised to $3500$, neighboring secondary filaments locally interact with one another, flatten into vortex sheets, and split into smaller tertiary filaments, as shown in Movie~\ref{mov:tubes_Re_3500} and Movie~\ref{mov:sheets}. The same interactions between secondary filaments occur for the Re$_{\Gamma} = 4500$ configuration; however, the secondary filaments become more disordered as they undergo complex 3D motion and become wrapped around each other and the original tubes, breaking down into fine-scale vortex filaments (see Movie~\ref{mov:tubes_Re_4500}). Lastly, in the Re$_{\Gamma} = 6000$ configuration, the secondary filaments rapidly emerge, interact, and violently break down as they almost instantaneously burst into an ensemble of vortices interacting over a wide range of scales, as shown in Movie~\ref{mov:tubes_Re_6000}. Notably, the high-circulation secondary filaments in this configuration locally interact to form new generations of perpendicular tertiary filaments, as shown in Fig. 5 in the main text and in Movie~\ref{mov:cascade}.
We propose that these tertiary filaments form through another iteration of the elliptical instability.

These results demonstrate how the elliptical instability provides a mechanism by which counter-rotating vortex tubes at high Reynolds numbers interact and break down to develop a turbulent cascade. This iterative instability generates new vortices that  interact with each other and ``grind down'' into smaller and smaller vortex filaments before being dissipated through viscosity~\cite{Taylor:1937}.  

\subsection{Vorticity evolution}
\label{subsec:Vorticity_Across_Re}
\begin{figure}[H]
\centering
\includegraphics[width=\textwidth]{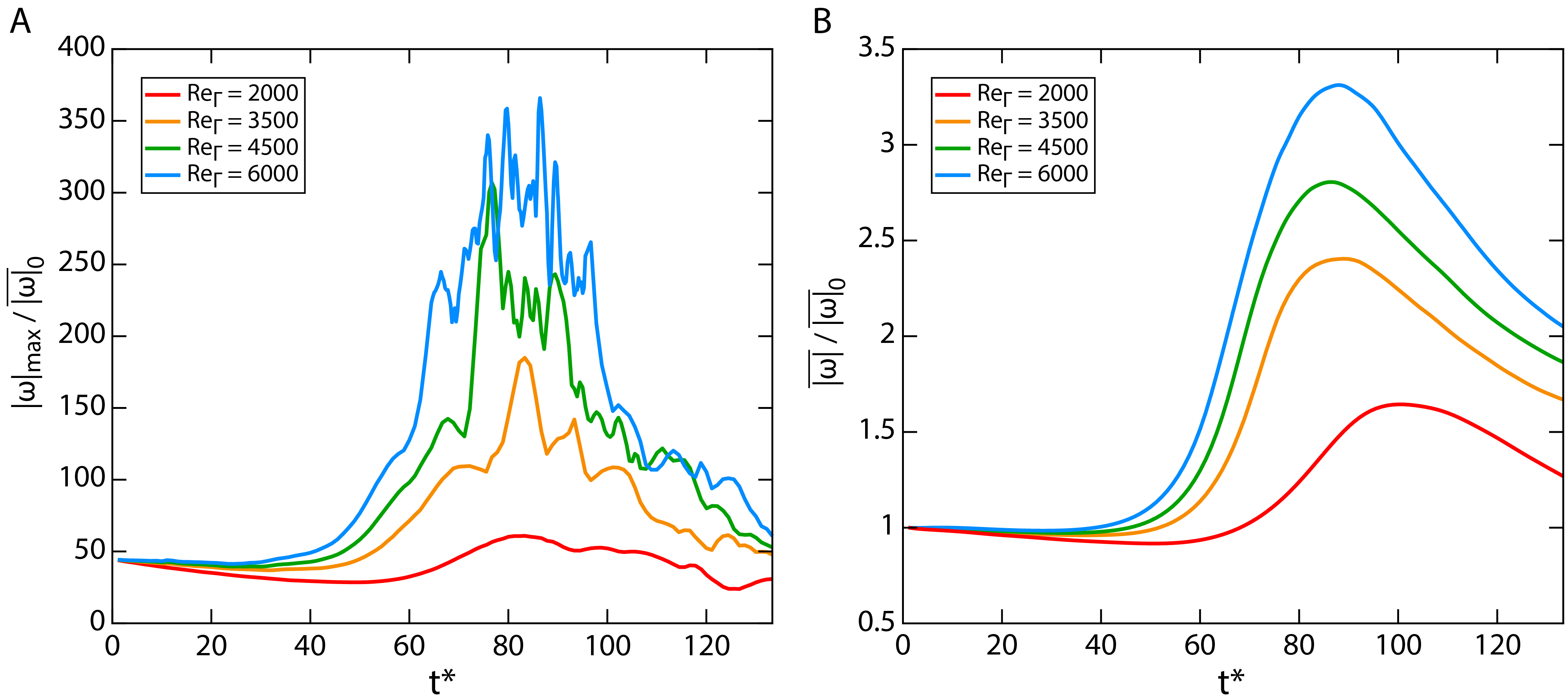}
\caption{Vorticity evolution for simulated vortex tube interactions at various Reynolds numbers. (A) Maximum vorticity modulus and (B) mean vorticity modulus evolution. Both moduli are normalized by the initial mean vorticity modulus. For each simulation, $\sigma = 0.06\mathcal{L}$, $b = 2.5\sigma$, Re$_{\Gamma} = \Gamma/\nu$, and $t^* = \Gamma t / b^2$. 
}
\label{fig:vorticity_evolution}
\end{figure}

The evolution of the vorticity modulus in the simulated vortex tube interactions also indicates the onset of a turbulent state during the breakdown of the tubes, which is especially pronounced for high-Reynolds number configurations. 
For each Reynolds number, the maximum vorticity modulus, $|\omega|_{\text{max}}$, remains initially constant until it increases during onset of the elliptical instability, as shown in Fig.~\ref{fig:vorticity_evolution}(A). 
For the Re$_{\Gamma} = 2000$ configuration, the maximum vorticity modulus increases slightly during the formation and stretching of the secondary filaments; however, because the filaments do not interact due to the onset of viscous dissipation, the maximum vorticity modulus decreases. 
For the higher-Reynolds number configurations, the maximum vorticity modulus increases precipitously due to the formation and stretching of the secondary filaments and remains sustained at a heightened level before decreasing due to viscous dissipation. 
This heightened level of $|\omega|_{\text{max}}$ coincides with the maximization of the energy dissipation rate, $\epsilon$, as shown in Fig.~\ref{fig:epsilon_energy_spectra}(A). 
The sustained amplification of the maximum vorticity modulus thus results from the new generation and local interactions of small-scale vortices during the turbulent breakdown. The higher the Reynolds number, the longer the maximum vorticity modulus remains at this elevated level before viscosity damps out the motion of the vortices.

Additionally, the mean vorticity modulus, $\overline{|\omega|}$, increases during the onset of the elliptical instability, reaches a peak value approximately when the dissipation rate, $\epsilon$, is maximized, and decreases as viscosity damps out the small-scale motion of the flow, as shown in Fig.~\ref{fig:vorticity_evolution}(B). 
The increase in the mean vorticity modulus indicates how the initially localized and coherent flow becomes more distributed throughout the domain during the breakdown. 
The amplification of the mean vorticity with increasing Reynolds number further demonstrates how the turbulent breakdown in the high-Reynolds number configurations is more three-dimensional and quasi-isotropic.   

\section{Supplemental movie descriptions}

\begin{movie}{Head-on collision of vortex rings. \\
 \normalfont{Underwater view of the head-on collision of two vortex rings dyed separately, where Re $ = UD/\nu = 6000$ and SR $= L/D = 2.5$. The rings expand radially as they collide at the midplane before rapidly breaking down into a turbulent cloud of dye.}
 }
\label{mov:GoPro}
\end{movie}
 
 \begin{movie}{Experimental vortex ring collision with dyed cores. \\
 \normalfont{Head-on view (top) and side view (bottom) showing the core dynamics of two colliding vortex rings, where Re $= 7000$ and SR $= 2.0$. As the rings stretch radially during the collision, the cores develop antisymmetric, short wavelength perturbations, indicative of the elliptical instability. Once fully formed, the perturbations deflect out-of-plane and break down into a turbulent cloud of dye.}
 }
 \label{mov:exp_cores}
 \end{movie}
 
 \begin{movie}{DNS of dyed vortex ring collision. \\
 \normalfont{Overhead view (top) and side view (bottom) showing the simulated collision of two vortex rings dyed red and blue, respectively (Re$_{\Gamma} = \Gamma/\nu = 4500$ and $t^{*} = \Gamma t / R_0^2$). The dye in the cores of the rings (dark) is differentiated from the dye surrounding the cores (light). As the rings collide, they stretch radially and develop antisymmetric, short-wavelength perturbations, indicative of the elliptical instability. As a result of these perturbations, the vortex rings interdigitate, forming alternating pairs of secondary vortex filaments, perpendicular to the cores. The rings then rapidly break down into a turbulent cloud of dye.}
 }
 \label{mov:sim_rings_dye}
 \end{movie}
 
 \begin{movie}{Experimental fully dyed vortex ring collision. \\
\normalfont{Overhead view (top) and side view (bottom) showing the collision of two vortex rings, where Re $= 6000$ and SR $= 2.5$. As the vortex rings collide, they develop alternating ``tongues'' that interdigitate around one another. The edges of these tongues roll up into an ordered array of secondary vortex filaments, perpendicular to the vortex cores. These secondary filaments interact and rapidly break down into a turbulent cloud of dye.}
}
\label{mov:exp_rings_dye}
\end{movie}

\begin{movie}{DNS of vortex tube interaction: Re$_{\Gamma} = 4500 $.   \\
\normalfont{Vorticity modulus for the simulated interaction of two antiparallel vortex tubes, where Re$_{\Gamma} = 4500$ and $t^{*} = \Gamma t / b^2$. As a result of the elliptical instability, the cores develop antisymmetric perturbations, and an array of counter-rotating secondary vortex filaments forms perpendicular to the cores. The secondary filaments interact with each other and the remains of the cores before breaking down into a ``soup'' of small-scale vortices that are dissipated by viscosity. The vorticity threshold is $0.076 \leq |\omega|/|\omega|_{\text{max}} \leq 0.198$, and the tubes propagate in the $-y$ direction. }
}
\label{mov:tubes_Re_4500}
\end{movie}

\begin{movie}{DNS of vortex tube interaction: Re$_{\Gamma} = 3500 $. \\
\normalfont{Vorticity modulus for the simulated interaction of two antiparallel vortex tubes, where Re$_{\Gamma} = 3500$ and $t^{*} = \Gamma t / b^2$. As a result of the elliptical instability, the cores develop antisymmetric perturbations, and an array of counter-rotating secondary vortex filaments forms perpendicular to the cores. The secondary filaments interact with each other and the remains of the cores before breaking down into a ``soup'' of small-scale vortices that are dissipated by viscosity. The vorticity threshold is $0.122 \leq |\omega|/|\omega|_{\text{max}} \leq 0.275$, and the tubes propagate in the $-y$ direction. }
}
\label{mov:tubes_Re_3500}
\end{movie}

\begin{movie}{Interaction and splitting of secondary vortex filaments. \\
\normalfont{Vorticity modulus for the simulated interaction of two antiparallel vortex tubes, where Re$_{\Gamma} = 3500$ and  $t^{*} = \Gamma t / b^2$. Neighboring secondary filaments counter-rotate and interact with one another. This close-range interaction causes one of the filaments to become flattened into a vortex sheet before splitting into smaller tertiary vortex filaments. The vorticity threshold is $0.122 \leq |\omega|/|\omega|_{\text{max}} \leq 0.206$. }
}
\label{mov:sheets}
\end{movie}

\begin{movie}{DNS of vortex tube interaction: Re$_{\Gamma} = 6000$. \\
\normalfont{Vorticity modulus for the simulated interaction of two antiparallel vortex tubes, where Re$_{\Gamma} = 6000$ and $t^{*} = \Gamma t / b^2$. As a result of the elliptical instability, the cores develop antisymmetric perturbations, and an array of counter-rotating secondary vortex filaments forms perpendicular to the cores. The secondary filaments and remaining cores interact violently and rapidly ``burst'' into a turbulent flow of vortices interacting over many scales. Viscosity damps out the motion of the remaining vortices. The vorticity threshold is $0.077 \leq |\omega|/|\omega|_{\text{max}} \leq 0.153$, and the tubes propagate in the $-y$ direction.}
}
\label{mov:tubes_Re_6000}
\end{movie}

\begin{movie}{Iterative cascade of elliptical instabilities. \\
\normalfont{Vorticity modulus for the simulated interaction of two antiparallel vortex tubes, where Re$_{\Gamma} = 6000$ and $t^{*} = \Gamma t / b^2$. Neighboring secondary filaments violently interact and form another generation of perpendicular, tertiary filaments through the elliptical instability. The vorticity threshold is $0.092 \leq |\omega|/|\omega|_{\text{max}} \leq 0.214$.}
}
\label{mov:cascade}
\end{movie}

\begin{movie}{Gaussian fit to vortex core PIV data. \\
\normalfont{2D PIV measurement of the vorticity distribution of a formed vortex ring, where Re $= 7000$ and SR $= 2.5$. The raw PIV vorticity data is plotted in the top panel and the Gaussian fit of the top and bottom cores is plotted in the bottom panel.}
}
\label{mov:PIV}
\end{movie}

\begin{movie}{DNS of vortex tube interaction: Re $_{\Gamma} = 2000$. \\
\normalfont{Vorticity modulus for the simulated interaction of two antiparallel vortex tubes, where Re$_{\Gamma} = 2000$ and $t^{*} = \Gamma t / b^2$. As a result of the elliptical instability, the cores develop antisymmetric perturbations, and an array of counter-rotating secondary vortex filaments forms perpendicular to the cores. The secondary filaments have little circulation and quickly dissipate due to viscosity. The vorticity threshold is $0.229 \leq |\omega|/|\omega|_{\text{max}} \leq 0.458$, and the tubes propagate in the $-y$ direction.}
}
\label{mov:tubes_Re_2000}
\end{movie}

\bibliography{supplemental}